\documentclass[superscriptaddress,reprint,amsmath,amssymb,aps,pra,longbibliography]{revtex4-1}

\usepackage{dcolumn} 

\setcitestyle{numbers,super}
\usepackage{placeins} 
\usepackage{upgreek}
\usepackage{amsmath} 
\usepackage{physics} 
\usepackage{graphicx} 
\usepackage{bm} \usepackage[margin=0.7in]{geometry}
\usepackage{natbib}

\begin{document}
\title{CMOS-Integrated Diamond Nitrogen-Vacancy Quantum Sensor}

\author{Donggyu Kim}
\thanks{These authors contributed equally}
\affiliation{Research Laboratory of Electronics, Massachusetts Institute of Technology, 50 Vassar St, Cambridge, MA, 02139, USA}
\affiliation{Department of Mechanical Engineering, Massachusetts Institute of Technology, 77 Mass Ave, Cambridge, MA, 02139, USA}

\author{Mohamed I. Ibrahim}
\thanks{These authors contributed equally}
\affiliation{Microsystems Technology Laboratories, Massachusetts Institute of Technology, 60 Vassar Street, Cambridge, MA, 02139, USA}
\affiliation{Department of Electrical Engineering and Computer Science, 
Massachusetts Institute of Technology, 50 Vassar St, Cambridge, MA, 02139, USA}

\author{Christopher Foy}
\thanks{These authors contributed equally}
\affiliation{Research Laboratory of Electronics, Massachusetts Institute of Technology, 50 Vassar St, Cambridge, MA, 02139, USA}
\affiliation{Department of Electrical Engineering and Computer Science, Massachusetts Institute of Technology, 50 Vassar St, Cambridge, MA, 02139, USA}

\author{Matthew E. Trusheim}
\affiliation{Research Laboratory of Electronics, Massachusetts Institute of Technology, 50 Vassar St, Cambridge, MA, 02139, USA}
\affiliation{Department of Electrical Engineering and Computer Science, Massachusetts Institute of Technology, 50 Vassar St, Cambridge, MA, 02139, USA}

\author{Ruonan Han}
\email{ruonan@mit.edu}
\affiliation{Microsystems Technology Laboratories, Massachusetts Institute of Technology, 60 Vassar Street, Cambridge, MA, 02139, USA}
\affiliation{Department of Electrical Engineering and Computer Science, Massachusetts Institute of Technology, 50 Vassar St, Cambridge, MA, 02139, USA}

\author{Dirk R. Englund}
\email{englund@mit.edu}
\affiliation{Research Laboratory of Electronics, Massachusetts Institute of Technology, 50 Vassar St, Cambridge, MA, 02139, USA}
\affiliation{Microsystems Technology Laboratories, Massachusetts Institute of Technology, 60 Vassar Street, Cambridge, MA, 02139, USA}
\affiliation{Department of Electrical Engineering and Computer Science, Massachusetts Institute of Technology, 50 Vassar St, Cambridge, MA, 02139, USA}

\begin{abstract}
The nitrogen vacancy (NV) center in diamond has emerged as a leading solid-state quantum sensor for applications including magnetometry, electrometry, thermometry, and chemical sensing. However, an outstanding challenge for practical applications is that existing NV-based sensing techniques require bulky and discrete instruments for spin control and detection. Here, we address this challenge by integrating NV based quantum sensing with complementary metal-oxide-semiconductor (CMOS) technology. Through tailored CMOS-integrated microwave generation and photodetection, this work dramatically reduces the instrumentation footprint for quantum magnetometry and thermometry. This hybrid diamond-CMOS integration enables an ultra-compact and scalable platform for quantum sensing and quantum information processing. 
\end{abstract}

\maketitle

\begin{figure*}[!t]
\centering
\includegraphics[scale=0.35]{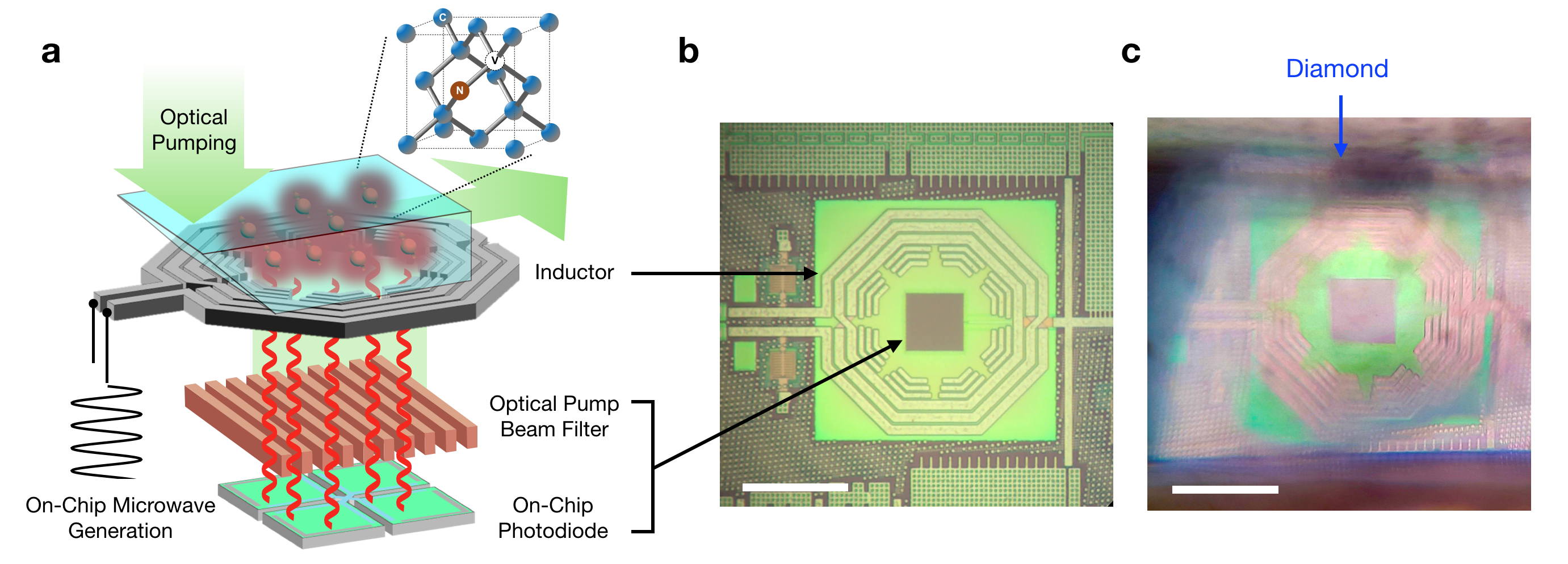}
\caption{\textbf{CMOS-integrated quantum sensing architecture.} 
\textbf{a,} A green pump laser excites an NV ensemble in the diamond slab. Microwave fields generated on-chip manipulate NV electron spins through an on-chip inductor, leading to ODMR. A metal/dielectric grating absorbs the green pump beam and transmits the NV spin-dependent fluorescence to the on-chip photodiode. Inset: NV atomic structure.
Top-view micrograph of the fabricated CMOS chip without (\textbf{b}) and with (\textbf{c}) the diamond slab. Scale bar is 200 $\upmu$m. 
}
\end{figure*}

Quantum metrology based on solid-state spins has shown outstanding sensing capabilities for various environmental physical quantities. In particular, the NV center in diamond has emerged as a leading room-temperature quantum sensor for temperature\cite{kucsko2013nanometre, neumann2013high, plakhotnik2014all, laraoui2015imaging}, strain\cite{ovartchaiyapong2014dynamic, teissier2014strain,trusheim2016wide}, electric fields\cite{dolde2011electric, chen2017high, Broadway:2018aa}, and magnetic fields\cite{maze2008nanoscale,balasubramanian2008nanoscale,grinolds2014subnanometre,jensen2014cavity,wolf2015subpicotesla, glenn2015single, boss2017quantum} especially to determine atomic species \cite{mamin2013nanoscale,staudacher2013nuclear,haberle2015nanoscale,rugar2015proton,aslam2017nanoscale,lovchinsky2016nuclear,lovchinsky2017magnetic,glenn2018high}. The advances of NV-based quantum metrology are based on its long spin coherence time\cite{balasubramanian2009ultralong} and its efficient optical interface for spin polarization and readout. Furthermore, picotesla magnetic field sensitivity at DC under ambient conditions has been achieved\cite{clevenson2015broadband} by interrogating NV center ensembles. This hybrid NV-CMOS platform is a highly advanced, scalable and compact platform for quantum sensing and will serve as a foundation for a new class of quantum systems.

Conventional approaches for NV magnetometry based on optically detected magnetic resonance (ODMR)\cite{taylor2008high} involve discrete off-the-shelf instruments that limit practical applications and scalability. NV ODMR requires (i) a microwave signal generator, amplifier, and delivery interface for NV spin manipulation, (ii) an optical filter to reject the pump laser, (iii) a photodetector for NV spin-dependent fluorescence measurement, and (iv) a pump laser. The use in conventional quantum sensing experiments of bulky instruments makes NV magnetometry hard outside of the lab or within mobile devices. 

Here, we realize a custom CMOS architecture that integrates requirements (i-iii) directly with a diamond sensor. This architecture\cite{Ibrahim2018VLSI} stacks the microwave inductor, filter, and photodiode into a 200 $\upmu$m $\times$ 200 $\upmu$m footprint. We use this hybrid diamond-CMOS platform to demonstrate ambient quantum magnetometry and thermometry. 

\section*{Chip-Scale Quantum Sensing}
\begin{figure}[!b]
\centering
\includegraphics[scale=0.30]{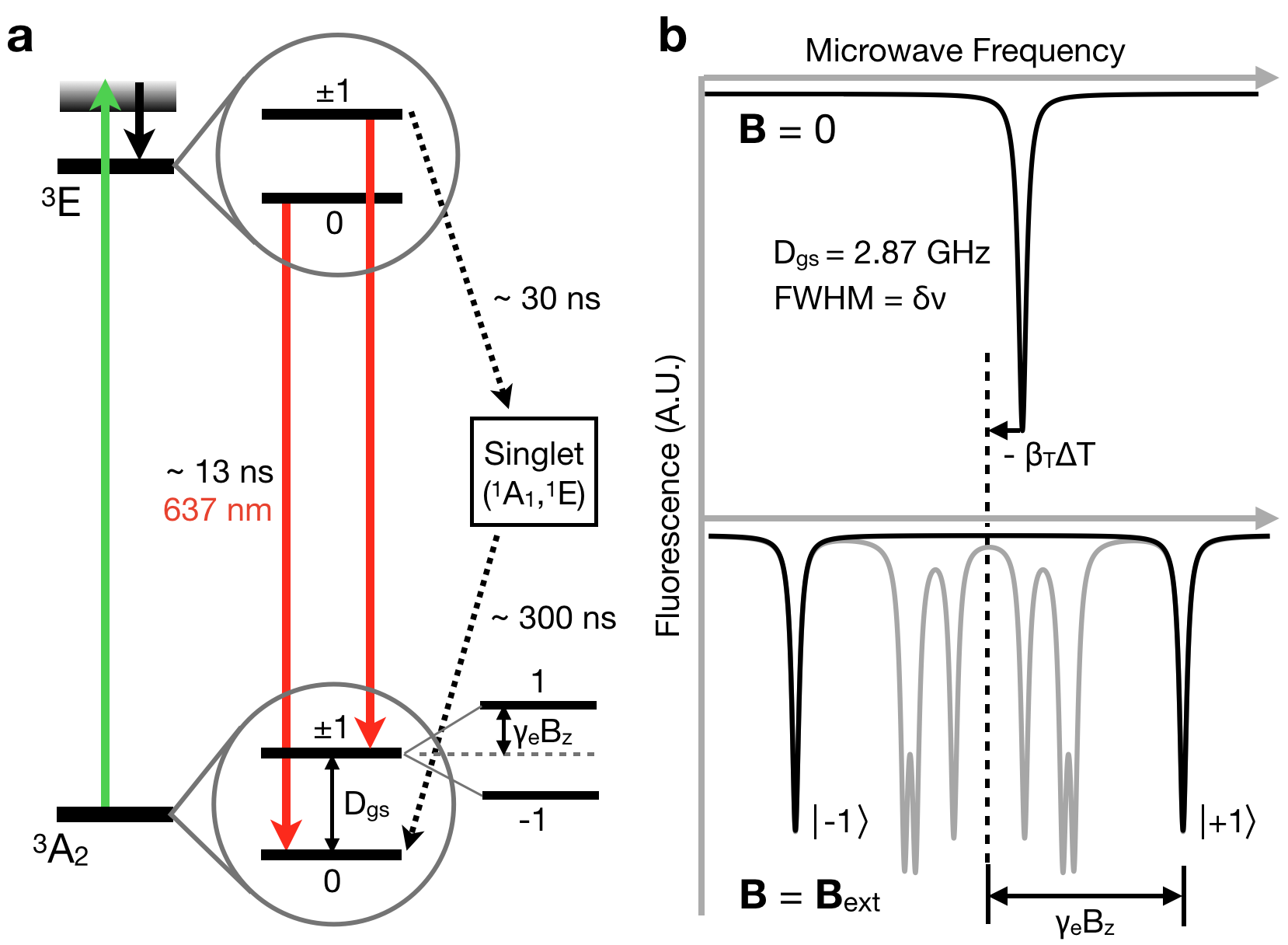}
\caption{\textbf{NV energy level diagram and ODMR spectra.} 
\textbf{a,} NV energy level diagram: green optical pump (green arrow) excites NV electrons from $^3\text{A}_2$ to $^3\text{E}$.  NV centers then emit red fluorescence by radiative decay (red arrow). Intersystem crossing rate (black dashed arrow) depends on NV spin states, resulting in spin-dependent fluorescence. 
\textbf{b,} Top: For $\textbf{B} = 0$, the ODMR spectrum shows one fluorescence dip for $\nu_{\pm} = D_\text{gs}$. Bottom: For an external magnetic field $\textbf{B}_\text{ext}$, this resonance splits into two Zeeman-shifted spin transitions (black curve), whose difference (mean) gives $B_z$ ($\Delta T$). The gray curves show ODMR for the other three possible NV orientations. 
}
\end{figure}

Figure 1a illustrates the device for on-chip ODMR. A diamond slab is irradiated and annealed to produce NV centers at a density of $\sim 0.01\text{ ppm}$. A 45$^{\circ}$ cut in the corner of the diamond directs the off-chip green pump beam along the length of the diamond slab. This side-excitation reduces pump laser background into the photodetector located below the diamond. An on-chip microwave generator and inductor drives the NV electron spin transitions. 

NV magnetometry detects external magnetic fields through the Zeeman shift induced on the NV's spin ground state sublevels\cite{taylor2008high}, shown in Fig. 2a. Specifically, an external magnetic field $\mathbf{B}$ induces an energy shift $\gamma_e B_z$ on the NV ground state spin triplet ($\ket{m_s=0,\pm 1}$), where $B_z$ is the magnetic field component along the NV symmetry axis. The spin transition frequencies, $\nu_{\pm}$, between sublevels $\ket{0}$ and $\ket{\pm1}$, are given by
\begin{equation}
\nu_{\pm} = (D_\text{gs} - \beta_{T} \Delta T) \pm \gamma_e B_z,
\end{equation} 
where $D_\text{gs} = 2.87 \text{ GHz}$ is the room-temperature natural ground-state splitting between sublevels $\ket{0}$ and $\ket{\pm1}$, $\gamma_e$ is the electronic gyromagnetic ratio ($28 \text{ GHz/\text{T}}$), $\beta_{T} = 74 \text{ kHz/K}$ \cite{acosta2010temperature}, and $\Delta T$ is the temperature shift from room temperature. Measuring $\nu_{\pm}$ gives $B_z$ and $\Delta T$ in their difference and sum, respectively. In addition, measuring $B_z$ for at least three of the four possible NV orientations in diamond (Inset in Fig. 1a) quantifies all components of $\mathbf{B}$ for vector magnetometry\cite{maertz2010vector,wang2015highvector, clevenson2018robust, schloss2018simultaneous}.

The NV ground state transitions $\nu_{\pm}$ are measured by ODMR under green laser excitation, as illustrated in Fig. 2a.  The spin magnetic sublevel $\ket{0}$ has a bright cycling transition, where it emits red fluorescence. In contrast, the $\ket{\pm 1}$ can undergo an intersystem crossing into a metastable, dark spin-singlet state, from where it decays back into the $\ket{0}$ sublevel. This has two consequences: optical spin polarization into sublevel $\ket{0}$ and lower average fluorescence of the $\ket{\pm 1}$ spin populations. The microwave field moves spin population between $\ket{0}$ and $\ket{\pm 1}$. Sweeping the applied microwave frequency leads to the ODMR spectra in Fig. 2b, from which $\nu_{\pm}$ are determined.

\section*{On-Chip microwave Generation and Delivery}

\begin{figure*}[t!]
\centering
\includegraphics[scale=0.35]{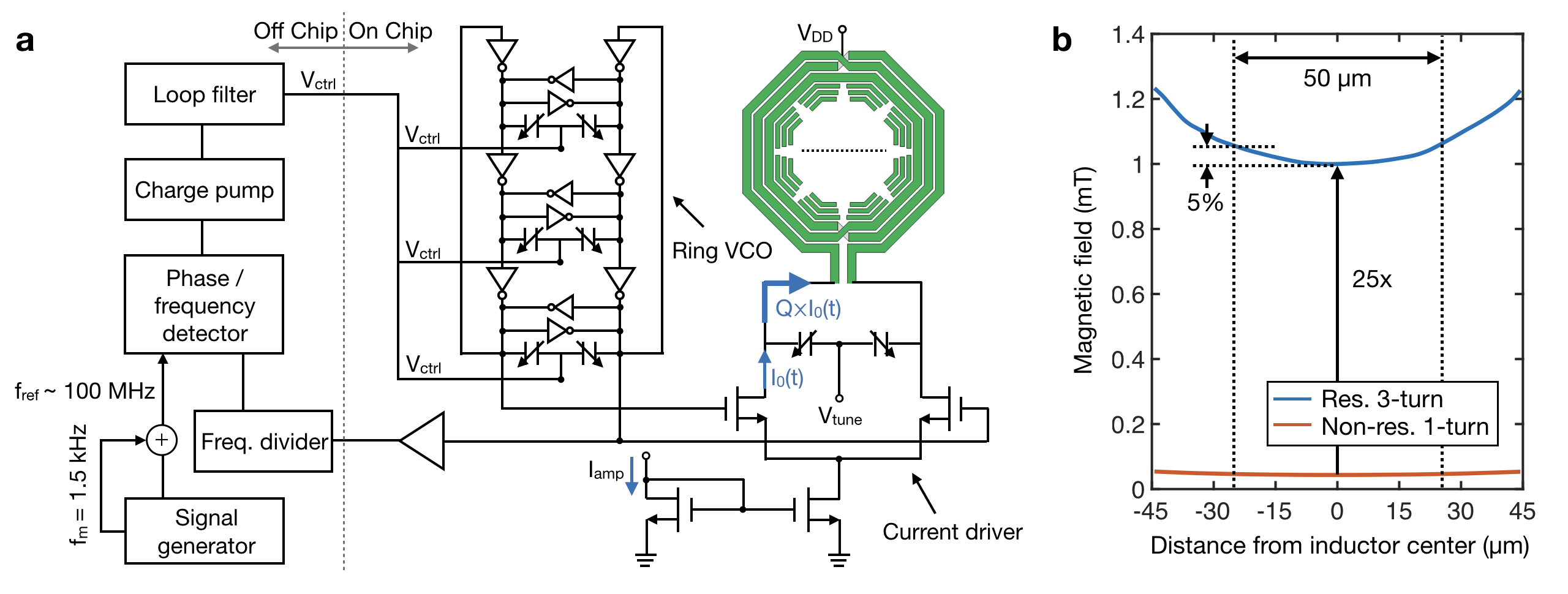}
\caption{\textbf{On-chip CMOS microwave generation and inductor characteristics.}
\textbf{a,} Schematic of microwave generation circuitry. 
\textbf{b,} High-frequency electromagnetic fields simulations (HFFS) for on-chip inductor: Magnetic field amplitude is plotted as a function of distance from the inductor center (dashed line in \textbf{a}). The resonant multi-turn loop inductor (blue) produces 25$\times$ higher amplitude compared to the nonresonant single turn inductor (red) at the same DC current. The insertion of the parasitic capacitive loops yields a microwave uniformity to 95\% over 50 $\upmu$m. 
}
\end{figure*}

\begin{figure}[!b]
\centering
\includegraphics[scale=0.3]{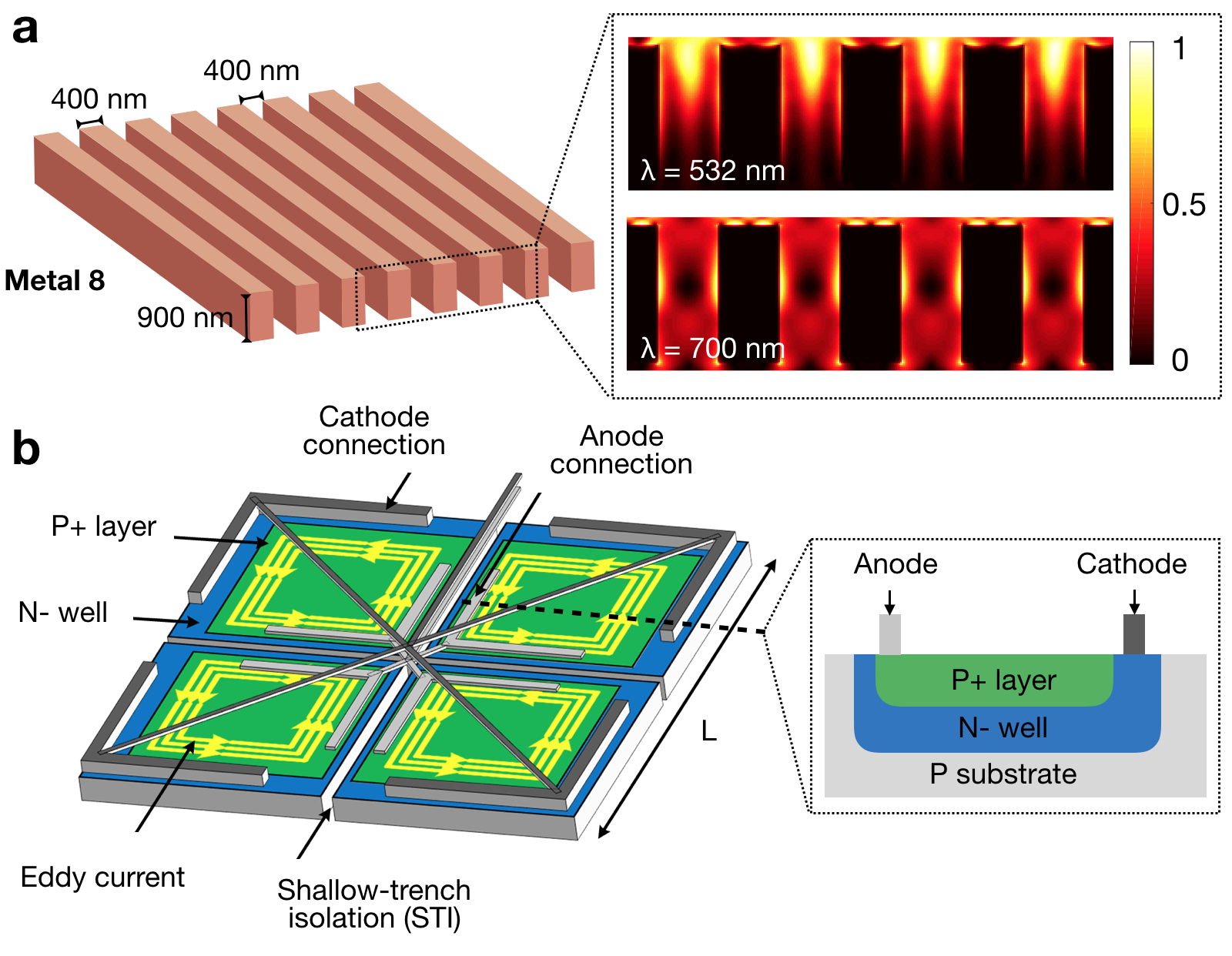}
\caption{\textbf{Optical detection of NV spin-dependent fluorescence.}
\textbf{a,} CMOS-compatible optical pump beam filter: the periodic metal-dielectric grating absorbs the green laser. Inset plots finite-difference-time-domain (FDTD) calculation of the optical intensity map inside the structure for green (top) and red light (bottom). Incident light polarization is perpendicular to the grating line.
\textbf{b,} Photodiode geometry: photodiode area is divided into four subareas, separated by a shallow trench isolation, to reduce eddy current (yellow loops) losses. Inset: cross-section along dashed line. 
}
\end{figure}

\begin{figure*}[!t]
\centering
\includegraphics[scale=0.35]{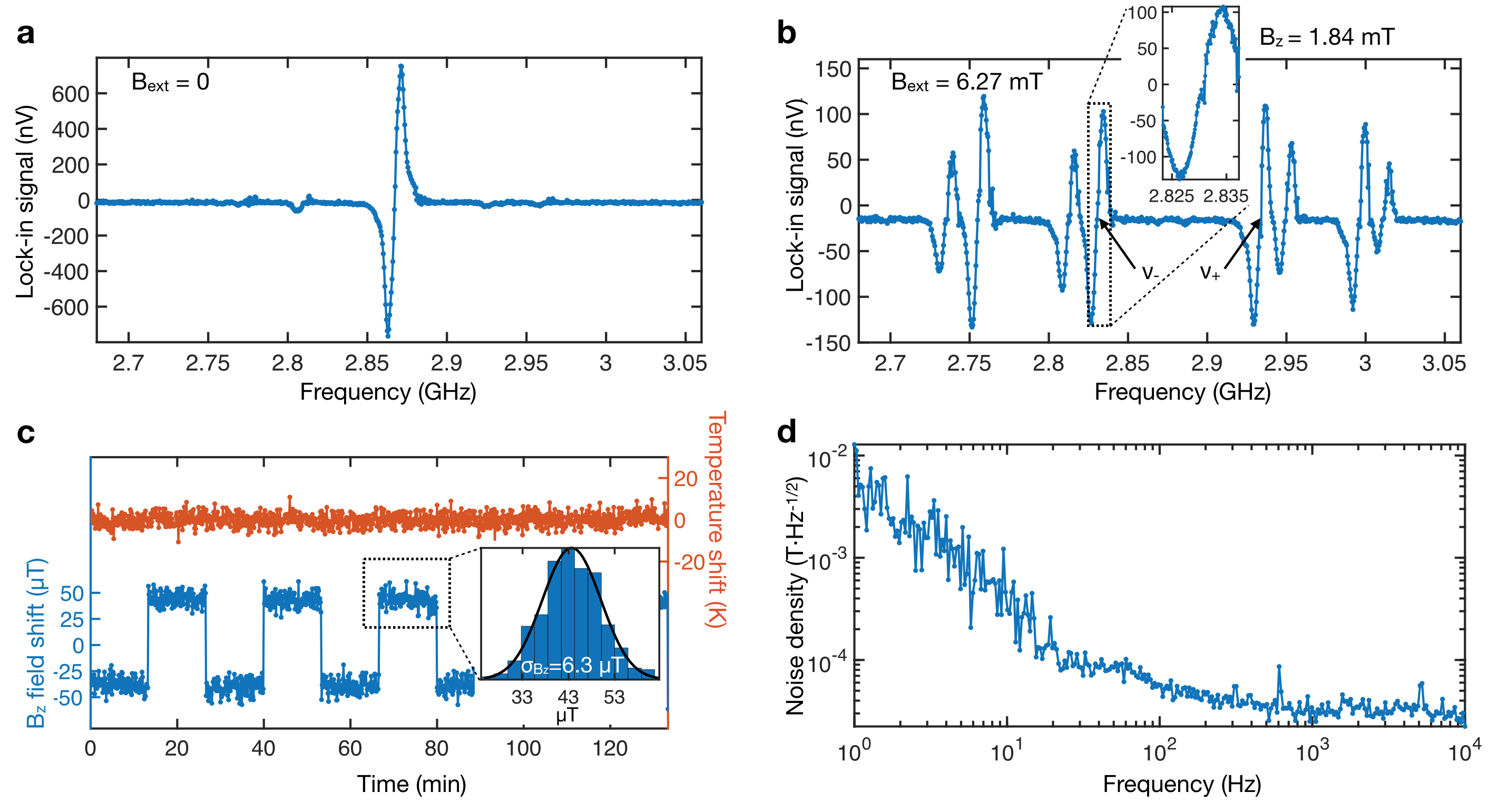}
\caption{\textbf{On-chip detection of ODMR and NV-based quantum magnetometry.} 
\textbf{a,} Frequency-modulated (FM) lock-in signal of NV spin-dependent fluorescence at zero external magnetic field  (in addition to $B \sim 100\  \upmu\text{T}$ of the earth magnetic field). 
\textbf{b,} FM lock-in signal with a permanent magnet ($B = 6.27 \text{mT}$): $B_z$ is the magnetic field along the NV-axis with the spin transition at $\nu_\pm$. The linewidth of the ODMR is 7 MHz. Slopes $d\text{V} / df$ at $\nu_- = 2.8303\text{ GHz}$ and $\nu_+ = 2.9330\text{ GHz}$ are $42.969 \text{ nV/MHz}$ and $42.450 \text{ nV/MHz}$, respectively.
\textbf{c,} On-chip magnetometry (Blue) and temperature effect (Red) separation: lock-in signals at both $\nu_{\pm}$ are observed while switching the polarity of external electromagnet with a period of 26 min. Inset plots the histogram of measured magnetic field $B_z$ with the standard deviation of 6.3 $\upmu$T. This uncertainty corresponds to a magnetic field sensitivity of 32.1 $\upmu$T$/\sqrt{\text{Hz}}$. The temperature sensitivity is 10.8 K$/\sqrt{\text{Hz}}$. Measurement is conducted with a time constant of 1 second.  
\textbf{d,} Noise spectral density monitored at $\nu_-$. 
} 
\end{figure*}

In our chip-scale NV magnetometer, the ground-state spin transitions are driven by the on-chip generated microwave fields. Figure 3a shows the circuitry for on-chip microwave generation and delivery. This circuitry is composed of a frequency synthesis loop, a current driver, and a resonant loop inductor. The frequency synthesis loop generates the microwave sweep signal from 2.6 GHz to 3.1 GHz required for the ODMR experiment. The main component of this loop is the on-chip CMOS ring voltage controlled oscillator (VCO). The ring VCO has a wide tuning range and avoids inductors\cite{razavi1998rf}, which minimizes the cross-talk between the oscillator and the microwave inductor which drives the NV ensemble. The loop is closed with off-chip components to enhance the stability and decrease the phase noise of the signal. 

The microwave fields are delivered to the NV ensemble through the loop inductor (Fig. 3a) implemented on the top most metal layer (Metal 9). To efficiently deliver microwave field, the loop inductor and a pair of shunt capacitors form a resonating load for the current driver. The load resonates near $D_\text{gs}$. This current driver is fed by the output of the ring VCO. Therefore, the current flowing in the inductor is at the microwave frequency. To improve the performance of this inductor for advanced NV sensing protocols we wish to increase the emitted MW field amplitude\cite{degen2017quantum}. The amplitude is enhanced by a factor Q of the driver DC bias current ($I_0 (t)$ $\sim 5 \text{ mA}$), where Q ($\sim 15$) is the quality factor of the inductor. In addition, we use a three-turn loop to multiply the microwave field strength. Overall, we have $25\times$ enhanced microwave field strength compared to a non-resonant single turn loop (as plotted in Fig. 3b). These protocols also require highly uniform microwave fields over the excitation volume. To achieve this, three capacitive parasitic loops are inserted\cite{Ibrahim2018VLSI}. We tailor the radius of the these loops, so that their opposite induced field homogenize the overall generated field. Another degree of freedom is the capacitive gaps in the parasitic loops. This controls the amount of current flowing in these loops. Therefore, we optimized these two parameters (i.e. the parasitic loop radius and the capacitive gap) for the three parasitic loops to achieve $>95\%$ uniformity.

\section*{On-chip Spin Readout}

The NV spin transitions are detected using an on-chip photodetector. A CMOS-compatible periodic metal-dielectric structure (Fig. 4a) in the Metal 8 interconnect layer filters green pump light. Specifically, incident light couples to the surfacepplasmon polariton (SPP) at the metal-dielectric interface, where green light rapidly decays due to frequency-dependent Ohmic loss\cite{zayats2005nano,hong2017fully}. The inset in Fig. 4a plots the intensity map for the green ($\lambda = 532 \text{ nm}$, Top) and red light ($\lambda = 700 \text{ nm}$, Bottom), showing the $\sim 95 \ \%$ and $\sim 5 \ \%$ absorption for green and red light through the structure, respectively. 

The photodiode consists of a P+/N-well/P-sub junctions (Inset in Fig. 4b), which is preferable for long wavelength detection\cite{murari2009photodiode}. Since we place the photodiode with its conductive layers below the inductor (Fig. 1a), large eddy currents near 2.87 GHz can be induced. This reduces the quality factor of the inductor, resulting in microwave amplitude reduction. We reduce this eddy current by half, by dividing the photodiode area into four subareas as shown in Fig. 4b (See Methods for detailed analysis). Furthermore, the anode/cathode connectors are arranged in a similar way to patterned ground shielding used in CMOS inductors\cite{yue1998chip}. This arrangement avoids any closed loops, which helps to cut the eddy current that may flow in the metallic connections. The photodiode has a measured responsivity of 0.23 A/W at the wavelength of 532 nm and a noise-equivalent-power of $\sim 4.9 \pm 2.6 \text{ nW}/\sqrt{\text{Hz}}$ at 1.5 kHz.

\section*{Experimental Results} 

We detect NV-ODMR with a lock-in technique. The green laser beam continuously excites the NV ensemble, and the frequency-modulated (FM) microwave fields ($f_m = 1.5 \text{ kHz}$ and modulation depth of 6 MHz) drive the NV electron spin transition. The spin-dependent fluorescence produces photo-current within the on-chip photodiode (Fig. 4b). Then, we read out the modulated photo-current at $f_m$ within an bandwidth of 0.078 Hz (a time constant of 1 second, considering the 24 dB/oct roll-off) with a lock-in amplifier (SR865A, Stanford Research System). 

Figure 5a shows the lock-in signal for the ODMR experiment under zero external magnetic field applied. This spectrum corresponds to the derivative of the ODMR spectrum shown in Fig. 2b. Next, we align a permanent magnet (6.27 mT) to split the spin transitions of the four NV orientations. Figure 5b shows the ODMR spectrum, which exhibits the expected eight spin transitions (Fig. 2b). In particular, we note the spin transitions at $\nu_- = 2.8303 \text{ GHz}$ and $\nu_+ = 2.9330 \text{ GHz}$ of the NV ensemble. 

Monitoring the lock-in signal $V$ at $\nu_-$ and $\nu_+$ allows independent measurements of magnetic field and temperature, as described above. Specifically, the sum of the lock-in signal change $\Delta V$ at $\nu_{\pm}$ is proportional to $\Delta T$, while the difference provides $\Delta B_\text{z}$:
\begin{equation}
\Delta T = \frac{1}{2\beta_T} \left( \eval{\frac{\Delta V}{dV / df}}_{\nu_+} + \eval{\frac{\Delta V}{dV / df}}_{\nu_-} \right)
\end{equation}
and
\begin{equation}
\Delta B_\text{z} = \frac{1}{2\gamma_e} \left( \eval{\frac{\Delta V}{dV / df}}_{\nu_+} - \eval{\frac{\Delta V}{dV / df}}_{\nu_-} \right).
\end{equation}
Figure 5c plots the detected $\Delta B_\text{z}$ induced by an electromagnet (blue) and measured $\Delta T$ (red).

The magnetic field sensitivity is given by the following relation:
\begin{equation}
S = \frac{\sigma_{B_z}}{\sqrt{\text{ENBW}}}.
\end{equation} 
Here, $\sigma_{B_z}$ is the noise in $\Delta B_\text{z}$ measurement, and ENBW is the equivalent noise bandwidth of the lock-in detector. In our measurement, $\text{ENBW} = 5 / (64\tau)$ with a time constant $\tau$ of 1 second, accounting for the $24 \ \text{dB/oct}$ of the lock-in amplifier roll-off. By measuring $\sigma_{B_z}$ of $6.3 \ \upmu\text{T}$ (Inset in Fig. 5c), we determine our DC magnetic field sensitivity of 32.1 $\upmu$T$/\sqrt{\text{Hz}}$ with a lock-in frequency of 1.5 kHz. Figure 5d plots the magnetic noise density measured at $\nu_{-}$ (no temperature compensation) with the noise floor of $\sim 27 \ \upmu$T$/\sqrt{\text{Hz}}$. 

\section*{Discussion}
The magnetic field sensitivity of our magnetometer is mainly limited by the green laser. Improving the on-chip optical filter performance by additional 60 dB (including nanophotonic structures\cite{peng1996resonant} in diamond or using multiple metal layers in CMOS) can improve the sensitivity by three orders of magnitude. In addition, recycling the green optical pump beam with a diamond wave-guide geometry\cite{clevenson2015broadband}, and implementing dynamical decoupling techniques\cite{taylor2008high, degen2017quantum} would also lead to orders of magnitude sensitivity improvements. 

In conclusion, we demonstrate chip-scale quantum magnetometry by integrating diamonds with CMOS technology. Throughout the CMOS multi-layers, essential components to detect NV ODMR - a microwave generator, an inductor, an optical pump beam filter, and a photodetector - are fabricated. NV spin ensembles integrated on the CMOS chip measure external magnetic fields with the sensitivity of 32.1 $\upmu$T/$\sqrt{\text{Hz}}$. This compact spin-CMOS platform can be extended toward on-chip sensing of other quantities such as electric fields. We emphasize that the CMOS circuit in this work provides direct physical interactions with the NV quantum states beyond electronic I/O signaling\cite{charbon201715}. 

In addition to chip-scale quantum sensing capability, our CMOS-based spin control and readout scheme will uniquely provide a scalable solution for implementing spin quantum-bit controls. This is in particular essential to develop a large-scale quantum information processor\cite{yao2012scalable, veldhorst2017silicon, charbon201715, patra2018cryo}, which enables quantum enhanced sensing\cite{giovannetti2011advances, unden2016quantum, degen2017quantum} and quantum information processing \cite{bernien2013heralded,pfaff2014unconditional,humphreys2018deterministic}. 

\section*{Methods}
\subsection*{Diamond and CMOS Preparation}
To avoid the direct injection of the green laser pump beam on the CMOS, we cut the CVD-grown diamond single crystal (Element 6) as shown in Fig. 1a, which enables the optical pumping in parallel with the CMOS-diamond interface with total internal reflection. This diamond is irradiated by the electron beam with a dosage of $10^{18}$ at 1 MeV. Then, the diamond is annealed for 2 hours at 850$^{\circ}$ Celsius. Our CMOS chip is fabricated with 65 nm LP TSMC technology. To reduce the background red fluorescence from the CMOS passivation layer, we etched the layer with plasma reactive-ion.

\subsection*{Measurement Setup}
A linearly polarized DPSS green laser beam (500 mW, $\lambda= 532\text{ nm}$, Verdi G2, Coherent) is delivered to the diamond through a telescope of $f_1 = 35\text{ mm}$ and $f_2 = 150\text{ mm}$. The beam diameter incident on the diamond is $\sim 500\text{ micron}$. A half-wave plate rotates the polarization of the laser beam to maximize the laser absorption through the periodic metal/dielectric structure in the Metal 8 layer. A permanent magnetic was used in Fig. 5b to split the NV orientations. The square-wave magnetic field applied in Fig. 5c was generated by an electromagnet. Alternating electrical current was used to avoid magnetization. 

\subsection*{Eddy Current Analysis}
For a square photodiode with a side length of L, the eddy current power $P_\text{eddy}$ is quadratically proportional to the change of the magnetic flux $d\phi(t)/dt$: 
\begin{equation*}
P_\text{eddy} \propto \frac{(d\phi(t)/dt)^2}{R} \propto \frac{L^4\left(\frac{dB}{dt}\right)^2}{L} \propto L^3 \left(\frac{dB}{dt}\right)^2.
\end{equation*}
Here, $t$ is the time; $R$ is the resistance; and $B$ is the magnetic field generated by the loop inductor in Metal 9. By dividing the photodiode active area into $N$ by $N$ subareas, the eddy current is reduced by $N^2 \times (L/N)^3 / L^3= 1/N$.

\section*{Acknowledgements}
This research is supported in part by the Army Research Office Multidisciplinary University Research Initiative (ARO MURI) biological transduction program. D.K. acknowledges financial support from the Kwanjeong Educational Foundation. M.I. acknowledges support from Singaporean-MIT Research Alliance (SMART), and MIT Center of Integrated Circuits and Systems. C.F. acknowledges support from Master Dynamic Limited. M.T. acknowledges support by an appointment to the Intelligence Community Postdoctoral Research Fellowship Program at MIT, administered by Oak Ridge Institute for Science and Education through an interagency agreement between the U.S. Department of Energy and the Office of the Director of National Intelligence. 

\section*{Author Contributions}
D.E. and R.H. initially conceived the diamond-CMOS integration. M.I. conceived the idea of stacking the microwave inductor, plasmonic filter, and photodiode in a 3D architecture.  M.I., C.F., and D.K. contributed to chip specifications, design and the experiment. M.I. constructed the CMOS chip prototype. D.K. performed FDTD simulation for the optical filter design and the diamond transfer on the CMOS chip. C.F. prepared the control software for the experiment. C.F. and D.K. constructed the optical setup and etched CMOS passivation layers. All authors contributed to the discussion of the experimental results and writing the manuscript.

\section*{Competing Interests}
The chip-scale spin control and detection scheme in this work was filed as a United States Provisional Patent Application (62/623151).

\section*{Materials and Correspondence}
The correspondence should be addressed to Ruonan Han (ruonan@mit.edu) or Dirk R. Englund (englund@mit.edu)


\begin{thebibliography}{50}%
\makeatletter
\providecommand \@ifxundefined [1]{%
 \@ifx{#1\undefined}
}%
\providecommand \@ifnum [1]{%
 \ifnum #1\expandafter \@firstoftwo
 \else \expandafter \@secondoftwo
 \fi
}%
\providecommand \@ifx [1]{%
 \ifx #1\expandafter \@firstoftwo
 \else \expandafter \@secondoftwo
 \fi
}%
\providecommand \natexlab [1]{#1}%
\providecommand \enquote  [1]{``#1''}%
\providecommand \bibnamefont  [1]{#1}%
\providecommand \bibfnamefont [1]{#1}%
\providecommand \citenamefont [1]{#1}%
\providecommand \href@noop [0]{\@secondoftwo}%
\providecommand \href [0]{\begingroup \@sanitize@url \@href}%
\providecommand \@href[1]{\@@startlink{#1}\@@href}%
\providecommand \@@href[1]{\endgroup#1\@@endlink}%
\providecommand \@sanitize@url [0]{\catcode `\\12\catcode `\$12\catcode
  `\&12\catcode `\#12\catcode `\^12\catcode `\_12\catcode `\%12\relax}%
\providecommand \@@startlink[1]{}%
\providecommand \@@endlink[0]{}%
\providecommand \url  [0]{\begingroup\@sanitize@url \@url }%
\providecommand \@url [1]{\endgroup\@href {#1}{\urlprefix }}%
\providecommand \urlprefix  [0]{URL }%
\providecommand \Eprint [0]{\href }%
\providecommand \doibase [0]{http://dx.doi.org/}%
\providecommand \selectlanguage [0]{\@gobble}%
\providecommand \bibinfo  [0]{\@secondoftwo}%
\providecommand \bibfield  [0]{\@secondoftwo}%
\providecommand \translation [1]{[#1]}%
\providecommand \BibitemOpen [0]{}%
\providecommand \bibitemStop [0]{}%
\providecommand \bibitemNoStop [0]{.\EOS\space}%
\providecommand \EOS [0]{\spacefactor3000\relax}%
\providecommand \BibitemShut  [1]{\csname bibitem#1\endcsname}%
\let\auto@bib@innerbib\@empty
%</preamble>
\bibitem [{\citenamefont {Kucsko}\ \emph {et~al.}(2013)\citenamefont {Kucsko},
  \citenamefont {Maurer}, \citenamefont {Yao}, \citenamefont {Kubo},
  \citenamefont {Noh}, \citenamefont {Lo}, \citenamefont {Park},\ and\
  \citenamefont {Lukin}}]{kucsko2013nanometre}%
  \BibitemOpen
  \bibfield  {author} {\bibinfo {author} {\bibfnamefont {Georg}\ \bibnamefont
  {Kucsko}}, \bibinfo {author} {\bibfnamefont {PC}~\bibnamefont {Maurer}},
  \bibinfo {author} {\bibfnamefont {Norman~Ying}\ \bibnamefont {Yao}}, \bibinfo
  {author} {\bibfnamefont {Michael}\ \bibnamefont {Kubo}}, \bibinfo {author}
  {\bibfnamefont {HJ}~\bibnamefont {Noh}}, \bibinfo {author} {\bibfnamefont
  {PK}~\bibnamefont {Lo}}, \bibinfo {author} {\bibfnamefont {Hongkun}\
  \bibnamefont {Park}}, \ and\ \bibinfo {author} {\bibfnamefont {Mikhail~D}\
  \bibnamefont {Lukin}},\ }\bibfield  {title} {\enquote {\bibinfo {title}
  {Nanometre-scale thermometry in a living cell},}\ }\href@noop {} {\bibfield
  {journal} {\bibinfo  {journal} {Nature}\ }\textbf {\bibinfo {volume} {500}},\
  \bibinfo {pages} {54} (\bibinfo {year} {2013})}\BibitemShut {NoStop}%
\bibitem [{\citenamefont {Neumann}\ \emph {et~al.}(2013)\citenamefont
  {Neumann}, \citenamefont {Jakobi}, \citenamefont {Dolde}, \citenamefont
  {Burk}, \citenamefont {Reuter}, \citenamefont {Waldherr}, \citenamefont
  {Honert}, \citenamefont {Wolf}, \citenamefont {Brunner}, \citenamefont {Shim}
  \emph {et~al.}}]{neumann2013high}%
  \BibitemOpen
  \bibfield  {author} {\bibinfo {author} {\bibfnamefont {Philipp}\ \bibnamefont
  {Neumann}}, \bibinfo {author} {\bibfnamefont {Ingmar}\ \bibnamefont
  {Jakobi}}, \bibinfo {author} {\bibfnamefont {Florian}\ \bibnamefont {Dolde}},
  \bibinfo {author} {\bibfnamefont {Christian}\ \bibnamefont {Burk}}, \bibinfo
  {author} {\bibfnamefont {Rolf}\ \bibnamefont {Reuter}}, \bibinfo {author}
  {\bibfnamefont {Gerald}\ \bibnamefont {Waldherr}}, \bibinfo {author}
  {\bibfnamefont {Jan}\ \bibnamefont {Honert}}, \bibinfo {author}
  {\bibfnamefont {Thomas}\ \bibnamefont {Wolf}}, \bibinfo {author}
  {\bibfnamefont {Andreas}\ \bibnamefont {Brunner}}, \bibinfo {author}
  {\bibfnamefont {Jeong~Hyun}\ \bibnamefont {Shim}},  \emph {et~al.},\
  }\bibfield  {title} {\enquote {\bibinfo {title} {High-precision nanoscale
  temperature sensing using single defects in diamond},}\ }\href@noop {}
  {\bibfield  {journal} {\bibinfo  {journal} {Nano letters}\ }\textbf {\bibinfo
  {volume} {13}},\ \bibinfo {pages} {2738--2742} (\bibinfo {year}
  {2013})}\BibitemShut {NoStop}%
\bibitem [{\citenamefont {Plakhotnik}\ \emph {et~al.}(2014)\citenamefont
  {Plakhotnik}, \citenamefont {Doherty}, \citenamefont {Cole}, \citenamefont
  {Chapman},\ and\ \citenamefont {Manson}}]{plakhotnik2014all}%
  \BibitemOpen
  \bibfield  {author} {\bibinfo {author} {\bibfnamefont {Taras}\ \bibnamefont
  {Plakhotnik}}, \bibinfo {author} {\bibfnamefont {Marcus~W}\ \bibnamefont
  {Doherty}}, \bibinfo {author} {\bibfnamefont {Jared~H}\ \bibnamefont {Cole}},
  \bibinfo {author} {\bibfnamefont {Robert}\ \bibnamefont {Chapman}}, \ and\
  \bibinfo {author} {\bibfnamefont {Neil~B}\ \bibnamefont {Manson}},\
  }\bibfield  {title} {\enquote {\bibinfo {title} {All-optical thermometry and
  thermal properties of the optically detected spin resonances of the
  nv--center in nanodiamond},}\ }\href@noop {} {\bibfield  {journal} {\bibinfo
  {journal} {Nano letters}\ }\textbf {\bibinfo {volume} {14}},\ \bibinfo
  {pages} {4989--4996} (\bibinfo {year} {2014})}\BibitemShut {NoStop}%
\bibitem [{\citenamefont {Laraoui}\ \emph {et~al.}(2015)\citenamefont
  {Laraoui}, \citenamefont {Aycock-Rizzo}, \citenamefont {Gao}, \citenamefont
  {Lu}, \citenamefont {Riedo},\ and\ \citenamefont
  {Meriles}}]{laraoui2015imaging}%
  \BibitemOpen
  \bibfield  {author} {\bibinfo {author} {\bibfnamefont {Abdelghani}\
  \bibnamefont {Laraoui}}, \bibinfo {author} {\bibfnamefont {Halley}\
  \bibnamefont {Aycock-Rizzo}}, \bibinfo {author} {\bibfnamefont {Yang}\
  \bibnamefont {Gao}}, \bibinfo {author} {\bibfnamefont {Xi}~\bibnamefont
  {Lu}}, \bibinfo {author} {\bibfnamefont {Elisa}\ \bibnamefont {Riedo}}, \
  and\ \bibinfo {author} {\bibfnamefont {Carlos~A}\ \bibnamefont {Meriles}},\
  }\bibfield  {title} {\enquote {\bibinfo {title} {Imaging thermal conductivity
  with nanoscale resolution using a scanning spin probe},}\ }\href@noop {}
  {\bibfield  {journal} {\bibinfo  {journal} {Nature communications}\ }\textbf
  {\bibinfo {volume} {6}},\ \bibinfo {pages} {8954} (\bibinfo {year}
  {2015})}\BibitemShut {NoStop}%
\bibitem [{\citenamefont {Ovartchaiyapong}\ \emph {et~al.}(2014)\citenamefont
  {Ovartchaiyapong}, \citenamefont {Lee}, \citenamefont {Myers},\ and\
  \citenamefont {Jayich}}]{ovartchaiyapong2014dynamic}%
  \BibitemOpen
  \bibfield  {author} {\bibinfo {author} {\bibfnamefont {Preeti}\ \bibnamefont
  {Ovartchaiyapong}}, \bibinfo {author} {\bibfnamefont {Kenneth~W}\
  \bibnamefont {Lee}}, \bibinfo {author} {\bibfnamefont {Bryan~A}\ \bibnamefont
  {Myers}}, \ and\ \bibinfo {author} {\bibfnamefont {Ania C~Bleszynski}\
  \bibnamefont {Jayich}},\ }\bibfield  {title} {\enquote {\bibinfo {title}
  {Dynamic strain-mediated coupling of a single diamond spin to a mechanical
  resonator},}\ }\href@noop {} {\bibfield  {journal} {\bibinfo  {journal}
  {Nature communications}\ }\textbf {\bibinfo {volume} {5}},\ \bibinfo {pages}
  {4429} (\bibinfo {year} {2014})}\BibitemShut {NoStop}%
\bibitem [{\citenamefont {Teissier}\ \emph {et~al.}(2014)\citenamefont
  {Teissier}, \citenamefont {Barfuss}, \citenamefont {Appel}, \citenamefont
  {Neu},\ and\ \citenamefont {Maletinsky}}]{teissier2014strain}%
  \BibitemOpen
  \bibfield  {author} {\bibinfo {author} {\bibfnamefont {J}~\bibnamefont
  {Teissier}}, \bibinfo {author} {\bibfnamefont {A}~\bibnamefont {Barfuss}},
  \bibinfo {author} {\bibfnamefont {P}~\bibnamefont {Appel}}, \bibinfo {author}
  {\bibfnamefont {E}~\bibnamefont {Neu}}, \ and\ \bibinfo {author}
  {\bibfnamefont {P}~\bibnamefont {Maletinsky}},\ }\bibfield  {title} {\enquote
  {\bibinfo {title} {Strain coupling of a nitrogen-vacancy center spin to a
  diamond mechanical oscillator},}\ }\href@noop {} {\bibfield  {journal}
  {\bibinfo  {journal} {Physical review letters}\ }\textbf {\bibinfo {volume}
  {113}},\ \bibinfo {pages} {020503} (\bibinfo {year} {2014})}\BibitemShut
  {NoStop}%
\bibitem [{\citenamefont {Trusheim}\ and\ \citenamefont
  {Englund}(2016)}]{trusheim2016wide}%
  \BibitemOpen
  \bibfield  {author} {\bibinfo {author} {\bibfnamefont {Matthew~E}\
  \bibnamefont {Trusheim}}\ and\ \bibinfo {author} {\bibfnamefont {Dirk}\
  \bibnamefont {Englund}},\ }\bibfield  {title} {\enquote {\bibinfo {title}
  {Wide-field strain imaging with preferentially aligned nitrogen-vacancy
  centers in polycrystalline diamond},}\ }\href@noop {} {\bibfield  {journal}
  {\bibinfo  {journal} {New Journal of Physics}\ }\textbf {\bibinfo {volume}
  {18}},\ \bibinfo {pages} {123023} (\bibinfo {year} {2016})}\BibitemShut
  {NoStop}%
\bibitem [{\citenamefont {Dolde}\ \emph {et~al.}(2011)\citenamefont {Dolde},
  \citenamefont {Fedder}, \citenamefont {Doherty}, \citenamefont {N{\"o}bauer},
  \citenamefont {Rempp}, \citenamefont {Balasubramanian}, \citenamefont {Wolf},
  \citenamefont {Reinhard}, \citenamefont {Hollenberg}, \citenamefont {Jelezko}
  \emph {et~al.}}]{dolde2011electric}%
  \BibitemOpen
  \bibfield  {author} {\bibinfo {author} {\bibfnamefont {Florian}\ \bibnamefont
  {Dolde}}, \bibinfo {author} {\bibfnamefont {Helmut}\ \bibnamefont {Fedder}},
  \bibinfo {author} {\bibfnamefont {Marcus~W}\ \bibnamefont {Doherty}},
  \bibinfo {author} {\bibfnamefont {Tobias}\ \bibnamefont {N{\"o}bauer}},
  \bibinfo {author} {\bibfnamefont {Florian}\ \bibnamefont {Rempp}}, \bibinfo
  {author} {\bibfnamefont {Gopalakrishnan}\ \bibnamefont {Balasubramanian}},
  \bibinfo {author} {\bibfnamefont {Thomas}\ \bibnamefont {Wolf}}, \bibinfo
  {author} {\bibfnamefont {Friedemann}\ \bibnamefont {Reinhard}}, \bibinfo
  {author} {\bibfnamefont {Lloyd~CL}\ \bibnamefont {Hollenberg}}, \bibinfo
  {author} {\bibfnamefont {Fedor}\ \bibnamefont {Jelezko}},  \emph {et~al.},\
  }\bibfield  {title} {\enquote {\bibinfo {title} {Electric-field sensing using
  single diamond spins},}\ }\href@noop {} {\bibfield  {journal} {\bibinfo
  {journal} {Nature Physics}\ }\textbf {\bibinfo {volume} {7}},\ \bibinfo
  {pages} {459} (\bibinfo {year} {2011})}\BibitemShut {NoStop}%
\bibitem [{\citenamefont {Chen}\ \emph {et~al.}(2017)\citenamefont {Chen},
  \citenamefont {Clevenson}, \citenamefont {Johnson}, \citenamefont {Pham},
  \citenamefont {Englund}, \citenamefont {Hemmer},\ and\ \citenamefont
  {Braje}}]{chen2017high}%
  \BibitemOpen
  \bibfield  {author} {\bibinfo {author} {\bibfnamefont {Edward~H}\
  \bibnamefont {Chen}}, \bibinfo {author} {\bibfnamefont {Hannah~A}\
  \bibnamefont {Clevenson}}, \bibinfo {author} {\bibfnamefont {Kerry~A}\
  \bibnamefont {Johnson}}, \bibinfo {author} {\bibfnamefont {Linh~M}\
  \bibnamefont {Pham}}, \bibinfo {author} {\bibfnamefont {Dirk~R}\ \bibnamefont
  {Englund}}, \bibinfo {author} {\bibfnamefont {Philip~R}\ \bibnamefont
  {Hemmer}}, \ and\ \bibinfo {author} {\bibfnamefont {Danielle~A}\ \bibnamefont
  {Braje}},\ }\bibfield  {title} {\enquote {\bibinfo {title} {High-sensitivity
  spin-based electrometry with an ensemble of nitrogen-vacancy centers in
  diamond},}\ }\href@noop {} {\bibfield  {journal} {\bibinfo  {journal}
  {Physical Review A}\ }\textbf {\bibinfo {volume} {95}},\ \bibinfo {pages}
  {053417} (\bibinfo {year} {2017})}\BibitemShut {NoStop}%
\bibitem [{\citenamefont {Broadway}\ \emph {et~al.}(2018)\citenamefont
  {Broadway}, \citenamefont {Dontschuk}, \citenamefont {Tsai}, \citenamefont
  {Lillie}, \citenamefont {Lew}, \citenamefont {McCallum}, \citenamefont
  {Johnson}, \citenamefont {Doherty}, \citenamefont {Stacey}, \citenamefont
  {Hollenberg},\ and\ \citenamefont {Tetienne}}]{Broadway:2018aa}%
  \BibitemOpen
  \bibfield  {author} {\bibinfo {author} {\bibfnamefont {D.~A.}\ \bibnamefont
  {Broadway}}, \bibinfo {author} {\bibfnamefont {N.}~\bibnamefont {Dontschuk}},
  \bibinfo {author} {\bibfnamefont {A.}~\bibnamefont {Tsai}}, \bibinfo {author}
  {\bibfnamefont {S.~E.}\ \bibnamefont {Lillie}}, \bibinfo {author}
  {\bibfnamefont {C.~T.~K.}\ \bibnamefont {Lew}}, \bibinfo {author}
  {\bibfnamefont {J.~C.}\ \bibnamefont {McCallum}}, \bibinfo {author}
  {\bibfnamefont {B.~C.}\ \bibnamefont {Johnson}}, \bibinfo {author}
  {\bibfnamefont {M.~W.}\ \bibnamefont {Doherty}}, \bibinfo {author}
  {\bibfnamefont {A.}~\bibnamefont {Stacey}}, \bibinfo {author} {\bibfnamefont
  {L.~C.~L.}\ \bibnamefont {Hollenberg}}, \ and\ \bibinfo {author}
  {\bibfnamefont {J.~P.}\ \bibnamefont {Tetienne}},\ }\bibfield  {title}
  {\enquote {\bibinfo {title} {Spatial mapping of band bending in semiconductor
  devices using in situ quantum sensors},}\ }\href {\doibase
  10.1038/s41928-018-0130-0} {\bibfield  {journal} {\bibinfo  {journal} {Nature
  Electronics}\ }\textbf {\bibinfo {volume} {1}},\ \bibinfo {pages} {502--507}
  (\bibinfo {year} {2018})}\BibitemShut {NoStop}%
\bibitem [{\citenamefont {Maze}\ \emph {et~al.}(2008)\citenamefont {Maze},
  \citenamefont {Stanwix}, \citenamefont {Hodges}, \citenamefont {Hong},
  \citenamefont {Taylor}, \citenamefont {Cappellaro}, \citenamefont {Jiang},
  \citenamefont {Dutt}, \citenamefont {Togan}, \citenamefont {Zibrov} \emph
  {et~al.}}]{maze2008nanoscale}%
  \BibitemOpen
  \bibfield  {author} {\bibinfo {author} {\bibfnamefont {JR}~\bibnamefont
  {Maze}}, \bibinfo {author} {\bibfnamefont {PL}~\bibnamefont {Stanwix}},
  \bibinfo {author} {\bibfnamefont {JS}~\bibnamefont {Hodges}}, \bibinfo
  {author} {\bibfnamefont {S}~\bibnamefont {Hong}}, \bibinfo {author}
  {\bibfnamefont {JM}~\bibnamefont {Taylor}}, \bibinfo {author} {\bibfnamefont
  {P}~\bibnamefont {Cappellaro}}, \bibinfo {author} {\bibfnamefont
  {L}~\bibnamefont {Jiang}}, \bibinfo {author} {\bibfnamefont {MV~Gurudev}\
  \bibnamefont {Dutt}}, \bibinfo {author} {\bibfnamefont {E}~\bibnamefont
  {Togan}}, \bibinfo {author} {\bibfnamefont {AS}~\bibnamefont {Zibrov}},
  \emph {et~al.},\ }\bibfield  {title} {\enquote {\bibinfo {title} {Nanoscale
  magnetic sensing with an individual electronic spin in diamond},}\
  }\href@noop {} {\bibfield  {journal} {\bibinfo  {journal} {Nature}\ }\textbf
  {\bibinfo {volume} {455}},\ \bibinfo {pages} {644} (\bibinfo {year}
  {2008})}\BibitemShut {NoStop}%
\bibitem [{\citenamefont {Balasubramanian}\ \emph {et~al.}(2008)\citenamefont
  {Balasubramanian}, \citenamefont {Chan}, \citenamefont {Kolesov},
  \citenamefont {Al-Hmoud}, \citenamefont {Tisler}, \citenamefont {Shin},
  \citenamefont {Kim}, \citenamefont {Wojcik}, \citenamefont {Hemmer},
  \citenamefont {Krueger} \emph {et~al.}}]{balasubramanian2008nanoscale}%
  \BibitemOpen
  \bibfield  {author} {\bibinfo {author} {\bibfnamefont {Gopalakrishnan}\
  \bibnamefont {Balasubramanian}}, \bibinfo {author} {\bibfnamefont
  {IY}~\bibnamefont {Chan}}, \bibinfo {author} {\bibfnamefont {Roman}\
  \bibnamefont {Kolesov}}, \bibinfo {author} {\bibfnamefont {Mohannad}\
  \bibnamefont {Al-Hmoud}}, \bibinfo {author} {\bibfnamefont {Julia}\
  \bibnamefont {Tisler}}, \bibinfo {author} {\bibfnamefont {Chang}\
  \bibnamefont {Shin}}, \bibinfo {author} {\bibfnamefont {Changdong}\
  \bibnamefont {Kim}}, \bibinfo {author} {\bibfnamefont {Aleksander}\
  \bibnamefont {Wojcik}}, \bibinfo {author} {\bibfnamefont {Philip~R}\
  \bibnamefont {Hemmer}}, \bibinfo {author} {\bibfnamefont {Anke}\ \bibnamefont
  {Krueger}},  \emph {et~al.},\ }\bibfield  {title} {\enquote {\bibinfo {title}
  {Nanoscale imaging magnetometry with diamond spins under ambient
  conditions},}\ }\href@noop {} {\bibfield  {journal} {\bibinfo  {journal}
  {Nature}\ }\textbf {\bibinfo {volume} {455}},\ \bibinfo {pages} {648}
  (\bibinfo {year} {2008})}\BibitemShut {NoStop}%
\bibitem [{\citenamefont {Grinolds}\ \emph {et~al.}(2014)\citenamefont
  {Grinolds}, \citenamefont {Warner}, \citenamefont {De~Greve}, \citenamefont
  {Dovzhenko}, \citenamefont {Thiel}, \citenamefont {Walsworth}, \citenamefont
  {Hong}, \citenamefont {Maletinsky},\ and\ \citenamefont
  {Yacoby}}]{grinolds2014subnanometre}%
  \BibitemOpen
  \bibfield  {author} {\bibinfo {author} {\bibfnamefont {MS}~\bibnamefont
  {Grinolds}}, \bibinfo {author} {\bibfnamefont {M}~\bibnamefont {Warner}},
  \bibinfo {author} {\bibfnamefont {Kristiaan}\ \bibnamefont {De~Greve}},
  \bibinfo {author} {\bibfnamefont {Yuliya}\ \bibnamefont {Dovzhenko}},
  \bibinfo {author} {\bibfnamefont {L}~\bibnamefont {Thiel}}, \bibinfo {author}
  {\bibfnamefont {Ronald~Lee}\ \bibnamefont {Walsworth}}, \bibinfo {author}
  {\bibfnamefont {S}~\bibnamefont {Hong}}, \bibinfo {author} {\bibfnamefont
  {P}~\bibnamefont {Maletinsky}}, \ and\ \bibinfo {author} {\bibfnamefont
  {Amir}\ \bibnamefont {Yacoby}},\ }\bibfield  {title} {\enquote {\bibinfo
  {title} {Subnanometre resolution in three-dimensional magnetic resonance
  imaging of individual dark spins},}\ }\href@noop {} {\bibfield  {journal}
  {\bibinfo  {journal} {Nature nanotechnology}\ }\textbf {\bibinfo {volume}
  {9}},\ \bibinfo {pages} {279} (\bibinfo {year} {2014})}\BibitemShut {NoStop}%
\bibitem [{\citenamefont {Jensen}\ \emph {et~al.}(2014)\citenamefont {Jensen},
  \citenamefont {Leefer}, \citenamefont {Jarmola}, \citenamefont {Dumeige},
  \citenamefont {Acosta}, \citenamefont {Kehayias}, \citenamefont {Patton},\
  and\ \citenamefont {Budker}}]{jensen2014cavity}%
  \BibitemOpen
  \bibfield  {author} {\bibinfo {author} {\bibfnamefont {Kasper}\ \bibnamefont
  {Jensen}}, \bibinfo {author} {\bibfnamefont {Nathan}\ \bibnamefont {Leefer}},
  \bibinfo {author} {\bibfnamefont {Andrey}\ \bibnamefont {Jarmola}}, \bibinfo
  {author} {\bibfnamefont {Yannick}\ \bibnamefont {Dumeige}}, \bibinfo {author}
  {\bibfnamefont {Victor~M}\ \bibnamefont {Acosta}}, \bibinfo {author}
  {\bibfnamefont {Pauli}\ \bibnamefont {Kehayias}}, \bibinfo {author}
  {\bibfnamefont {Brian}\ \bibnamefont {Patton}}, \ and\ \bibinfo {author}
  {\bibfnamefont {Dmitry}\ \bibnamefont {Budker}},\ }\bibfield  {title}
  {\enquote {\bibinfo {title} {Cavity-enhanced room-temperature magnetometry
  using absorption by nitrogen-vacancy centers in diamond},}\ }\href@noop {}
  {\bibfield  {journal} {\bibinfo  {journal} {Physical review letters}\
  }\textbf {\bibinfo {volume} {112}},\ \bibinfo {pages} {160802} (\bibinfo
  {year} {2014})}\BibitemShut {NoStop}%
\bibitem [{\citenamefont {Wolf}\ \emph {et~al.}(2015)\citenamefont {Wolf},
  \citenamefont {Neumann}, \citenamefont {Nakamura}, \citenamefont {Sumiya},
  \citenamefont {Ohshima}, \citenamefont {Isoya},\ and\ \citenamefont
  {Wrachtrup}}]{wolf2015subpicotesla}%
  \BibitemOpen
  \bibfield  {author} {\bibinfo {author} {\bibfnamefont {Thomas}\ \bibnamefont
  {Wolf}}, \bibinfo {author} {\bibfnamefont {Philipp}\ \bibnamefont {Neumann}},
  \bibinfo {author} {\bibfnamefont {Kazuo}\ \bibnamefont {Nakamura}}, \bibinfo
  {author} {\bibfnamefont {Hitoshi}\ \bibnamefont {Sumiya}}, \bibinfo {author}
  {\bibfnamefont {Takeshi}\ \bibnamefont {Ohshima}}, \bibinfo {author}
  {\bibfnamefont {Junichi}\ \bibnamefont {Isoya}}, \ and\ \bibinfo {author}
  {\bibfnamefont {J{\"o}rg}\ \bibnamefont {Wrachtrup}},\ }\bibfield  {title}
  {\enquote {\bibinfo {title} {Subpicotesla diamond magnetometry},}\
  }\href@noop {} {\bibfield  {journal} {\bibinfo  {journal} {Physical Review
  X}\ }\textbf {\bibinfo {volume} {5}},\ \bibinfo {pages} {041001} (\bibinfo
  {year} {2015})}\BibitemShut {NoStop}%
\bibitem [{\citenamefont {Glenn}\ \emph {et~al.}(2015)\citenamefont {Glenn},
  \citenamefont {Lee}, \citenamefont {Park}, \citenamefont {Weissleder},
  \citenamefont {Yacoby}, \citenamefont {Lukin}, \citenamefont {Lee},
  \citenamefont {Walsworth},\ and\ \citenamefont {Connolly}}]{glenn2015single}%
  \BibitemOpen
  \bibfield  {author} {\bibinfo {author} {\bibfnamefont {David~R}\ \bibnamefont
  {Glenn}}, \bibinfo {author} {\bibfnamefont {Kyungheon}\ \bibnamefont {Lee}},
  \bibinfo {author} {\bibfnamefont {Hongkun}\ \bibnamefont {Park}}, \bibinfo
  {author} {\bibfnamefont {Ralph}\ \bibnamefont {Weissleder}}, \bibinfo
  {author} {\bibfnamefont {Amir}\ \bibnamefont {Yacoby}}, \bibinfo {author}
  {\bibfnamefont {Mikhail~D}\ \bibnamefont {Lukin}}, \bibinfo {author}
  {\bibfnamefont {Hakho}\ \bibnamefont {Lee}}, \bibinfo {author} {\bibfnamefont
  {Ronald~L}\ \bibnamefont {Walsworth}}, \ and\ \bibinfo {author}
  {\bibfnamefont {Colin~B}\ \bibnamefont {Connolly}},\ }\bibfield  {title}
  {\enquote {\bibinfo {title} {Single-cell magnetic imaging using a quantum
  diamond microscope},}\ }\href@noop {} {\bibfield  {journal} {\bibinfo
  {journal} {Nature methods}\ }\textbf {\bibinfo {volume} {12}},\ \bibinfo
  {pages} {736} (\bibinfo {year} {2015})}\BibitemShut {NoStop}%
\bibitem [{\citenamefont {Boss}\ \emph {et~al.}(2017)\citenamefont {Boss},
  \citenamefont {Cujia}, \citenamefont {Zopes},\ and\ \citenamefont
  {Degen}}]{boss2017quantum}%
  \BibitemOpen
  \bibfield  {author} {\bibinfo {author} {\bibfnamefont {Jens~M}\ \bibnamefont
  {Boss}}, \bibinfo {author} {\bibfnamefont {KS}~\bibnamefont {Cujia}},
  \bibinfo {author} {\bibfnamefont {Jonathan}\ \bibnamefont {Zopes}}, \ and\
  \bibinfo {author} {\bibfnamefont {Christian~L}\ \bibnamefont {Degen}},\
  }\bibfield  {title} {\enquote {\bibinfo {title} {Quantum sensing with
  arbitrary frequency resolution},}\ }\href@noop {} {\bibfield  {journal}
  {\bibinfo  {journal} {Science}\ }\textbf {\bibinfo {volume} {356}},\ \bibinfo
  {pages} {837--840} (\bibinfo {year} {2017})}\BibitemShut {NoStop}%
\bibitem [{\citenamefont {Mamin}\ \emph {et~al.}(2013)\citenamefont {Mamin},
  \citenamefont {Kim}, \citenamefont {Sherwood}, \citenamefont {Rettner},
  \citenamefont {Ohno}, \citenamefont {Awschalom},\ and\ \citenamefont
  {Rugar}}]{mamin2013nanoscale}%
  \BibitemOpen
  \bibfield  {author} {\bibinfo {author} {\bibfnamefont {HJ}~\bibnamefont
  {Mamin}}, \bibinfo {author} {\bibfnamefont {M}~\bibnamefont {Kim}}, \bibinfo
  {author} {\bibfnamefont {MH}~\bibnamefont {Sherwood}}, \bibinfo {author}
  {\bibfnamefont {CT}~\bibnamefont {Rettner}}, \bibinfo {author} {\bibfnamefont
  {K}~\bibnamefont {Ohno}}, \bibinfo {author} {\bibfnamefont {DD}~\bibnamefont
  {Awschalom}}, \ and\ \bibinfo {author} {\bibfnamefont {D}~\bibnamefont
  {Rugar}},\ }\bibfield  {title} {\enquote {\bibinfo {title} {Nanoscale nuclear
  magnetic resonance with a nitrogen-vacancy spin sensor},}\ }\href@noop {}
  {\bibfield  {journal} {\bibinfo  {journal} {Science}\ }\textbf {\bibinfo
  {volume} {339}},\ \bibinfo {pages} {557--560} (\bibinfo {year}
  {2013})}\BibitemShut {NoStop}%
\bibitem [{\citenamefont {Staudacher}\ \emph {et~al.}(2013)\citenamefont
  {Staudacher}, \citenamefont {Shi}, \citenamefont {Pezzagna}, \citenamefont
  {Meijer}, \citenamefont {Du}, \citenamefont {Meriles}, \citenamefont
  {Reinhard},\ and\ \citenamefont {Wrachtrup}}]{staudacher2013nuclear}%
  \BibitemOpen
  \bibfield  {author} {\bibinfo {author} {\bibfnamefont {Tobias}\ \bibnamefont
  {Staudacher}}, \bibinfo {author} {\bibfnamefont {Fazhan}\ \bibnamefont
  {Shi}}, \bibinfo {author} {\bibfnamefont {S}~\bibnamefont {Pezzagna}},
  \bibinfo {author} {\bibfnamefont {Jan}\ \bibnamefont {Meijer}}, \bibinfo
  {author} {\bibfnamefont {Jiangfeng}\ \bibnamefont {Du}}, \bibinfo {author}
  {\bibfnamefont {Carlos~A}\ \bibnamefont {Meriles}}, \bibinfo {author}
  {\bibfnamefont {Friedemann}\ \bibnamefont {Reinhard}}, \ and\ \bibinfo
  {author} {\bibfnamefont {Joerg}\ \bibnamefont {Wrachtrup}},\ }\bibfield
  {title} {\enquote {\bibinfo {title} {Nuclear magnetic resonance spectroscopy
  on a (5-nanometer) 3 sample volume},}\ }\href@noop {} {\bibfield  {journal}
  {\bibinfo  {journal} {Science}\ }\textbf {\bibinfo {volume} {339}},\ \bibinfo
  {pages} {561--563} (\bibinfo {year} {2013})}\BibitemShut {NoStop}%
\bibitem [{\citenamefont {H{\"a}berle}\ \emph {et~al.}(2015)\citenamefont
  {H{\"a}berle}, \citenamefont {Schmid-Lorch}, \citenamefont {Reinhard},\ and\
  \citenamefont {Wrachtrup}}]{haberle2015nanoscale}%
  \BibitemOpen
  \bibfield  {author} {\bibinfo {author} {\bibfnamefont {T}~\bibnamefont
  {H{\"a}berle}}, \bibinfo {author} {\bibfnamefont {D}~\bibnamefont
  {Schmid-Lorch}}, \bibinfo {author} {\bibfnamefont {F}~\bibnamefont
  {Reinhard}}, \ and\ \bibinfo {author} {\bibfnamefont {J}~\bibnamefont
  {Wrachtrup}},\ }\bibfield  {title} {\enquote {\bibinfo {title} {Nanoscale
  nuclear magnetic imaging with chemical contrast},}\ }\href@noop {} {\bibfield
   {journal} {\bibinfo  {journal} {Nature nanotechnology}\ }\textbf {\bibinfo
  {volume} {10}},\ \bibinfo {pages} {125} (\bibinfo {year} {2015})}\BibitemShut
  {NoStop}%
\bibitem [{\citenamefont {Rugar}\ \emph {et~al.}(2015)\citenamefont {Rugar},
  \citenamefont {Mamin}, \citenamefont {Sherwood}, \citenamefont {Kim},
  \citenamefont {Rettner}, \citenamefont {Ohno},\ and\ \citenamefont
  {Awschalom}}]{rugar2015proton}%
  \BibitemOpen
  \bibfield  {author} {\bibinfo {author} {\bibfnamefont {D}~\bibnamefont
  {Rugar}}, \bibinfo {author} {\bibfnamefont {HJ}~\bibnamefont {Mamin}},
  \bibinfo {author} {\bibfnamefont {MH}~\bibnamefont {Sherwood}}, \bibinfo
  {author} {\bibfnamefont {M}~\bibnamefont {Kim}}, \bibinfo {author}
  {\bibfnamefont {CT}~\bibnamefont {Rettner}}, \bibinfo {author} {\bibfnamefont
  {K}~\bibnamefont {Ohno}}, \ and\ \bibinfo {author} {\bibfnamefont
  {DD}~\bibnamefont {Awschalom}},\ }\bibfield  {title} {\enquote {\bibinfo
  {title} {Proton magnetic resonance imaging using a nitrogen--vacancy spin
  sensor},}\ }\href@noop {} {\bibfield  {journal} {\bibinfo  {journal} {Nature
  nanotechnology}\ }\textbf {\bibinfo {volume} {10}},\ \bibinfo {pages} {120}
  (\bibinfo {year} {2015})}\BibitemShut {NoStop}%
\bibitem [{\citenamefont {Aslam}\ \emph {et~al.}(2017)\citenamefont {Aslam},
  \citenamefont {Pfender}, \citenamefont {Neumann}, \citenamefont {Reuter},
  \citenamefont {Zappe}, \citenamefont {de~Oliveira}, \citenamefont
  {Denisenko}, \citenamefont {Sumiya}, \citenamefont {Onoda}, \citenamefont
  {Isoya} \emph {et~al.}}]{aslam2017nanoscale}%
  \BibitemOpen
  \bibfield  {author} {\bibinfo {author} {\bibfnamefont {Nabeel}\ \bibnamefont
  {Aslam}}, \bibinfo {author} {\bibfnamefont {Matthias}\ \bibnamefont
  {Pfender}}, \bibinfo {author} {\bibfnamefont {Philipp}\ \bibnamefont
  {Neumann}}, \bibinfo {author} {\bibfnamefont {Rolf}\ \bibnamefont {Reuter}},
  \bibinfo {author} {\bibfnamefont {Andrea}\ \bibnamefont {Zappe}}, \bibinfo
  {author} {\bibfnamefont {Felipe~F{\'a}varo}\ \bibnamefont {de~Oliveira}},
  \bibinfo {author} {\bibfnamefont {Andrej}\ \bibnamefont {Denisenko}},
  \bibinfo {author} {\bibfnamefont {Hitoshi}\ \bibnamefont {Sumiya}}, \bibinfo
  {author} {\bibfnamefont {Shinobu}\ \bibnamefont {Onoda}}, \bibinfo {author}
  {\bibfnamefont {Junichi}\ \bibnamefont {Isoya}},  \emph {et~al.},\ }\bibfield
   {title} {\enquote {\bibinfo {title} {Nanoscale nuclear magnetic resonance
  with chemical resolution},}\ }\href@noop {} {\bibfield  {journal} {\bibinfo
  {journal} {Science}\ }\textbf {\bibinfo {volume} {357}},\ \bibinfo {pages}
  {67--71} (\bibinfo {year} {2017})}\BibitemShut {NoStop}%
\bibitem [{\citenamefont {Lovchinsky}\ \emph {et~al.}(2016)\citenamefont
  {Lovchinsky}, \citenamefont {Sushkov}, \citenamefont {Urbach}, \citenamefont
  {de~Leon}, \citenamefont {Choi}, \citenamefont {De~Greve}, \citenamefont
  {Evans}, \citenamefont {Gertner}, \citenamefont {Bersin}, \citenamefont
  {M{\"u}ller} \emph {et~al.}}]{lovchinsky2016nuclear}%
  \BibitemOpen
  \bibfield  {author} {\bibinfo {author} {\bibfnamefont {Igor}\ \bibnamefont
  {Lovchinsky}}, \bibinfo {author} {\bibfnamefont {AO}~\bibnamefont {Sushkov}},
  \bibinfo {author} {\bibfnamefont {E}~\bibnamefont {Urbach}}, \bibinfo
  {author} {\bibfnamefont {NP}~\bibnamefont {de~Leon}}, \bibinfo {author}
  {\bibfnamefont {Soonwon}\ \bibnamefont {Choi}}, \bibinfo {author}
  {\bibfnamefont {Kristiaan}\ \bibnamefont {De~Greve}}, \bibinfo {author}
  {\bibfnamefont {R}~\bibnamefont {Evans}}, \bibinfo {author} {\bibfnamefont
  {R}~\bibnamefont {Gertner}}, \bibinfo {author} {\bibfnamefont
  {E}~\bibnamefont {Bersin}}, \bibinfo {author} {\bibfnamefont {C}~\bibnamefont
  {M{\"u}ller}},  \emph {et~al.},\ }\bibfield  {title} {\enquote {\bibinfo
  {title} {Nuclear magnetic resonance detection and spectroscopy of single
  proteins using quantum logic},}\ }\href@noop {} {\bibfield  {journal}
  {\bibinfo  {journal} {Science}\ }\textbf {\bibinfo {volume} {351}},\ \bibinfo
  {pages} {836--841} (\bibinfo {year} {2016})}\BibitemShut {NoStop}%
\bibitem [{\citenamefont {Lovchinsky}\ \emph {et~al.}(2017)\citenamefont
  {Lovchinsky}, \citenamefont {Sanchez-Yamagishi}, \citenamefont {Urbach},
  \citenamefont {Choi}, \citenamefont {Fang}, \citenamefont {Andersen},
  \citenamefont {Watanabe}, \citenamefont {Taniguchi}, \citenamefont
  {Bylinskii}, \citenamefont {Kaxiras} \emph
  {et~al.}}]{lovchinsky2017magnetic}%
  \BibitemOpen
  \bibfield  {author} {\bibinfo {author} {\bibfnamefont {Igor}\ \bibnamefont
  {Lovchinsky}}, \bibinfo {author} {\bibfnamefont {JD}~\bibnamefont
  {Sanchez-Yamagishi}}, \bibinfo {author} {\bibfnamefont {EK}~\bibnamefont
  {Urbach}}, \bibinfo {author} {\bibfnamefont {S}~\bibnamefont {Choi}},
  \bibinfo {author} {\bibfnamefont {S}~\bibnamefont {Fang}}, \bibinfo {author}
  {\bibfnamefont {TI}~\bibnamefont {Andersen}}, \bibinfo {author}
  {\bibfnamefont {K}~\bibnamefont {Watanabe}}, \bibinfo {author} {\bibfnamefont
  {T}~\bibnamefont {Taniguchi}}, \bibinfo {author} {\bibfnamefont
  {A}~\bibnamefont {Bylinskii}}, \bibinfo {author} {\bibfnamefont
  {E}~\bibnamefont {Kaxiras}},  \emph {et~al.},\ }\bibfield  {title} {\enquote
  {\bibinfo {title} {Magnetic resonance spectroscopy of an atomically thin
  material using a single-spin qubit},}\ }\href@noop {} {\bibfield  {journal}
  {\bibinfo  {journal} {Science}\ ,\ \bibinfo {pages} {eaal2538}} (\bibinfo
  {year} {2017})}\BibitemShut {NoStop}%
\bibitem [{\citenamefont {Glenn}\ \emph {et~al.}(2018)\citenamefont {Glenn},
  \citenamefont {Bucher}, \citenamefont {Lee}, \citenamefont {Lukin},
  \citenamefont {Park},\ and\ \citenamefont {Walsworth}}]{glenn2018high}%
  \BibitemOpen
  \bibfield  {author} {\bibinfo {author} {\bibfnamefont {David~R}\ \bibnamefont
  {Glenn}}, \bibinfo {author} {\bibfnamefont {Dominik~B}\ \bibnamefont
  {Bucher}}, \bibinfo {author} {\bibfnamefont {Junghyun}\ \bibnamefont {Lee}},
  \bibinfo {author} {\bibfnamefont {Mikhail~D}\ \bibnamefont {Lukin}}, \bibinfo
  {author} {\bibfnamefont {Hongkun}\ \bibnamefont {Park}}, \ and\ \bibinfo
  {author} {\bibfnamefont {Ronald~L}\ \bibnamefont {Walsworth}},\ }\bibfield
  {title} {\enquote {\bibinfo {title} {High-resolution magnetic resonance
  spectroscopy using a solid-state spin sensor},}\ }\href@noop {} {\bibfield
  {journal} {\bibinfo  {journal} {Nature}\ }\textbf {\bibinfo {volume} {555}},\
  \bibinfo {pages} {351} (\bibinfo {year} {2018})}\BibitemShut {NoStop}%
\bibitem [{\citenamefont {Balasubramanian}\ \emph {et~al.}(2009)\citenamefont
  {Balasubramanian}, \citenamefont {Neumann}, \citenamefont {Twitchen},
  \citenamefont {Markham}, \citenamefont {Kolesov}, \citenamefont {Mizuochi},
  \citenamefont {Isoya}, \citenamefont {Achard}, \citenamefont {Beck},
  \citenamefont {Tissler} \emph {et~al.}}]{balasubramanian2009ultralong}%
  \BibitemOpen
  \bibfield  {author} {\bibinfo {author} {\bibfnamefont {Gopalakrishnan}\
  \bibnamefont {Balasubramanian}}, \bibinfo {author} {\bibfnamefont {Philipp}\
  \bibnamefont {Neumann}}, \bibinfo {author} {\bibfnamefont {Daniel}\
  \bibnamefont {Twitchen}}, \bibinfo {author} {\bibfnamefont {Matthew}\
  \bibnamefont {Markham}}, \bibinfo {author} {\bibfnamefont {Roman}\
  \bibnamefont {Kolesov}}, \bibinfo {author} {\bibfnamefont {Norikazu}\
  \bibnamefont {Mizuochi}}, \bibinfo {author} {\bibfnamefont {Junichi}\
  \bibnamefont {Isoya}}, \bibinfo {author} {\bibfnamefont {Jocelyn}\
  \bibnamefont {Achard}}, \bibinfo {author} {\bibfnamefont {Johannes}\
  \bibnamefont {Beck}}, \bibinfo {author} {\bibfnamefont {Julia}\ \bibnamefont
  {Tissler}},  \emph {et~al.},\ }\bibfield  {title} {\enquote {\bibinfo {title}
  {Ultralong spin coherence time in isotopically engineered diamond},}\
  }\href@noop {} {\bibfield  {journal} {\bibinfo  {journal} {Nature materials}\
  }\textbf {\bibinfo {volume} {8}},\ \bibinfo {pages} {383} (\bibinfo {year}
  {2009})}\BibitemShut {NoStop}%
\bibitem [{\citenamefont {Clevenson}\ \emph {et~al.}(2015)\citenamefont
  {Clevenson}, \citenamefont {Trusheim}, \citenamefont {Teale}, \citenamefont
  {Schr{\"o}der}, \citenamefont {Braje},\ and\ \citenamefont
  {Englund}}]{clevenson2015broadband}%
  \BibitemOpen
  \bibfield  {author} {\bibinfo {author} {\bibfnamefont {Hannah}\ \bibnamefont
  {Clevenson}}, \bibinfo {author} {\bibfnamefont {Matthew~E}\ \bibnamefont
  {Trusheim}}, \bibinfo {author} {\bibfnamefont {Carson}\ \bibnamefont
  {Teale}}, \bibinfo {author} {\bibfnamefont {Tim}\ \bibnamefont
  {Schr{\"o}der}}, \bibinfo {author} {\bibfnamefont {Danielle}\ \bibnamefont
  {Braje}}, \ and\ \bibinfo {author} {\bibfnamefont {Dirk}\ \bibnamefont
  {Englund}},\ }\bibfield  {title} {\enquote {\bibinfo {title} {Broadband
  magnetometry and temperature sensing with a light-trapping diamond
  waveguide},}\ }\href@noop {} {\bibfield  {journal} {\bibinfo  {journal}
  {Nature Physics}\ }\textbf {\bibinfo {volume} {11}},\ \bibinfo {pages} {393}
  (\bibinfo {year} {2015})}\BibitemShut {NoStop}%
\bibitem [{\citenamefont {Taylor}\ \emph {et~al.}(2008)\citenamefont {Taylor},
  \citenamefont {Cappellaro}, \citenamefont {Childress}, \citenamefont {Jiang},
  \citenamefont {Budker}, \citenamefont {Hemmer}, \citenamefont {Yacoby},
  \citenamefont {Walsworth},\ and\ \citenamefont {Lukin}}]{taylor2008high}%
  \BibitemOpen
  \bibfield  {author} {\bibinfo {author} {\bibfnamefont {JM}~\bibnamefont
  {Taylor}}, \bibinfo {author} {\bibfnamefont {P}~\bibnamefont {Cappellaro}},
  \bibinfo {author} {\bibfnamefont {L}~\bibnamefont {Childress}}, \bibinfo
  {author} {\bibfnamefont {L}~\bibnamefont {Jiang}}, \bibinfo {author}
  {\bibfnamefont {D}~\bibnamefont {Budker}}, \bibinfo {author} {\bibfnamefont
  {PR}~\bibnamefont {Hemmer}}, \bibinfo {author} {\bibfnamefont
  {A}~\bibnamefont {Yacoby}}, \bibinfo {author} {\bibfnamefont {R}~\bibnamefont
  {Walsworth}}, \ and\ \bibinfo {author} {\bibfnamefont {MD}~\bibnamefont
  {Lukin}},\ }\bibfield  {title} {\enquote {\bibinfo {title} {High-sensitivity
  diamond magnetometer with nanoscale resolution},}\ }\href@noop {} {\bibfield
  {journal} {\bibinfo  {journal} {Nature Physics}\ }\textbf {\bibinfo {volume}
  {4}},\ \bibinfo {pages} {810} (\bibinfo {year} {2008})}\BibitemShut {NoStop}%
\bibitem [{\citenamefont {Ibrahim}\ \emph {et~al.}(2018)\citenamefont
  {Ibrahim}, \citenamefont {Foy}, \citenamefont {Kim}, \citenamefont
  {Englund},\ and\ \citenamefont {Han}}]{Ibrahim2018VLSI}%
  \BibitemOpen
  \bibfield  {author} {\bibinfo {author} {\bibfnamefont {Mohamed~I.}\
  \bibnamefont {Ibrahim}}, \bibinfo {author} {\bibfnamefont {Christopher}\
  \bibnamefont {Foy}}, \bibinfo {author} {\bibfnamefont {Donggyu}\ \bibnamefont
  {Kim}}, \bibinfo {author} {\bibfnamefont {Dirk~R.}\ \bibnamefont {Englund}},
  \ and\ \bibinfo {author} {\bibfnamefont {Ruonan}\ \bibnamefont {Han}},\
  }\bibfield  {title} {\enquote {\bibinfo {title} {Room-temperature quantum
  sensing in cmos: On-chip detection of electronic spin states in diamond color
  centers for magnetometry},}\ }\href@noop {} {\bibfield  {journal} {\bibinfo
  {journal} {IEEE VLSI Circuits Symposium}\ } (\bibinfo {year}
  {2018})}\BibitemShut {NoStop}%
\bibitem [{\citenamefont {Acosta}\ \emph {et~al.}(2010)\citenamefont {Acosta},
  \citenamefont {Bauch}, \citenamefont {Ledbetter}, \citenamefont {Waxman},
  \citenamefont {Bouchard},\ and\ \citenamefont
  {Budker}}]{acosta2010temperature}%
  \BibitemOpen
  \bibfield  {author} {\bibinfo {author} {\bibfnamefont {VM}~\bibnamefont
  {Acosta}}, \bibinfo {author} {\bibfnamefont {E}~\bibnamefont {Bauch}},
  \bibinfo {author} {\bibfnamefont {MP}~\bibnamefont {Ledbetter}}, \bibinfo
  {author} {\bibfnamefont {A}~\bibnamefont {Waxman}}, \bibinfo {author}
  {\bibfnamefont {L-S}\ \bibnamefont {Bouchard}}, \ and\ \bibinfo {author}
  {\bibfnamefont {D}~\bibnamefont {Budker}},\ }\bibfield  {title} {\enquote
  {\bibinfo {title} {Temperature dependence of the nitrogen-vacancy magnetic
  resonance in diamond},}\ }\href@noop {} {\bibfield  {journal} {\bibinfo
  {journal} {Physical review letters}\ }\textbf {\bibinfo {volume} {104}},\
  \bibinfo {pages} {070801} (\bibinfo {year} {2010})}\BibitemShut {NoStop}%
\bibitem [{\citenamefont {Maertz}\ \emph {et~al.}(2010)\citenamefont {Maertz},
  \citenamefont {Wijnheijmer}, \citenamefont {Fuchs}, \citenamefont
  {Nowakowski},\ and\ \citenamefont {Awschalom}}]{maertz2010vector}%
  \BibitemOpen
  \bibfield  {author} {\bibinfo {author} {\bibfnamefont {BJ}~\bibnamefont
  {Maertz}}, \bibinfo {author} {\bibfnamefont {AP}~\bibnamefont {Wijnheijmer}},
  \bibinfo {author} {\bibfnamefont {GD}~\bibnamefont {Fuchs}}, \bibinfo
  {author} {\bibfnamefont {ME}~\bibnamefont {Nowakowski}}, \ and\ \bibinfo
  {author} {\bibfnamefont {DD}~\bibnamefont {Awschalom}},\ }\bibfield  {title}
  {\enquote {\bibinfo {title} {Vector magnetic field microscopy using nitrogen
  vacancy centers in diamond},}\ }\href@noop {} {\bibfield  {journal} {\bibinfo
   {journal} {Applied Physics Letters}\ }\textbf {\bibinfo {volume} {96}},\
  \bibinfo {pages} {092504} (\bibinfo {year} {2010})}\BibitemShut {NoStop}%
\bibitem [{\citenamefont {Wang}\ \emph {et~al.}(2015)\citenamefont {Wang},
  \citenamefont {Yuan}, \citenamefont {Huang}, \citenamefont {Rong},
  \citenamefont {Wang}, \citenamefont {Xu}, \citenamefont {Duan}, \citenamefont
  {Ju}, \citenamefont {Shi},\ and\ \citenamefont {Du}}]{wang2015highvector}%
  \BibitemOpen
  \bibfield  {author} {\bibinfo {author} {\bibfnamefont {Pengfei}\ \bibnamefont
  {Wang}}, \bibinfo {author} {\bibfnamefont {Zhenheng}\ \bibnamefont {Yuan}},
  \bibinfo {author} {\bibfnamefont {Pu}~\bibnamefont {Huang}}, \bibinfo
  {author} {\bibfnamefont {Xing}\ \bibnamefont {Rong}}, \bibinfo {author}
  {\bibfnamefont {Mengqi}\ \bibnamefont {Wang}}, \bibinfo {author}
  {\bibfnamefont {Xiangkun}\ \bibnamefont {Xu}}, \bibinfo {author}
  {\bibfnamefont {Changkui}\ \bibnamefont {Duan}}, \bibinfo {author}
  {\bibfnamefont {Chenyong}\ \bibnamefont {Ju}}, \bibinfo {author}
  {\bibfnamefont {Fazhan}\ \bibnamefont {Shi}}, \ and\ \bibinfo {author}
  {\bibfnamefont {Jiangfeng}\ \bibnamefont {Du}},\ }\bibfield  {title}
  {\enquote {\bibinfo {title} {High-resolution vector microwave magnetometry
  based on solid-state spins in diamond},}\ }\href@noop {} {\bibfield
  {journal} {\bibinfo  {journal} {Nature communications}\ }\textbf {\bibinfo
  {volume} {6}},\ \bibinfo {pages} {6631} (\bibinfo {year} {2015})}\BibitemShut
  {NoStop}%
\bibitem [{\citenamefont {Clevenson}\ \emph {et~al.}(2018)\citenamefont
  {Clevenson}, \citenamefont {Pham}, \citenamefont {Teale}, \citenamefont
  {Johnson}, \citenamefont {Englund},\ and\ \citenamefont
  {Braje}}]{clevenson2018robust}%
  \BibitemOpen
  \bibfield  {author} {\bibinfo {author} {\bibfnamefont {Hannah}\ \bibnamefont
  {Clevenson}}, \bibinfo {author} {\bibfnamefont {Linh~M}\ \bibnamefont
  {Pham}}, \bibinfo {author} {\bibfnamefont {Carson}\ \bibnamefont {Teale}},
  \bibinfo {author} {\bibfnamefont {Kerry}\ \bibnamefont {Johnson}}, \bibinfo
  {author} {\bibfnamefont {Dirk}\ \bibnamefont {Englund}}, \ and\ \bibinfo
  {author} {\bibfnamefont {Danielle}\ \bibnamefont {Braje}},\ }\bibfield
  {title} {\enquote {\bibinfo {title} {Robust high-dynamic-range vector
  magnetometry with nitrogen-vacancy centers in diamond},}\ }\href@noop {}
  {\bibfield  {journal} {\bibinfo  {journal} {Applied Physics Letters}\
  }\textbf {\bibinfo {volume} {112}},\ \bibinfo {pages} {252406} (\bibinfo
  {year} {2018})}\BibitemShut {NoStop}%
\bibitem [{\citenamefont {Schloss}\ \emph {et~al.}(2018)\citenamefont
  {Schloss}, \citenamefont {Barry}, \citenamefont {Turner},\ and\ \citenamefont
  {Walsworth}}]{schloss2018simultaneous}%
  \BibitemOpen
  \bibfield  {author} {\bibinfo {author} {\bibfnamefont {Jennifer~M}\
  \bibnamefont {Schloss}}, \bibinfo {author} {\bibfnamefont {John~F}\
  \bibnamefont {Barry}}, \bibinfo {author} {\bibfnamefont {Matthew~J}\
  \bibnamefont {Turner}}, \ and\ \bibinfo {author} {\bibfnamefont {Ronald~L}\
  \bibnamefont {Walsworth}},\ }\bibfield  {title} {\enquote {\bibinfo {title}
  {Simultaneous broadband vector magnetometry using solid-state spins},}\
  }\href@noop {} {\bibfield  {journal} {\bibinfo  {journal} {arXiv preprint
  arXiv:1803.03718}\ } (\bibinfo {year} {2018})}\BibitemShut {NoStop}%
\bibitem [{\citenamefont {Razavi}(1998)}]{razavi1998rf}%
  \BibitemOpen
  \bibfield  {author} {\bibinfo {author} {\bibfnamefont {Behzad}\ \bibnamefont
  {Razavi}},\ }\href@noop {} {\emph {\bibinfo {title} {RF microelectronics}}},\
  Vol.~\bibinfo {volume} {2}\ (\bibinfo  {publisher} {Prentice Hall New
  Jersey},\ \bibinfo {year} {1998})\BibitemShut {NoStop}%
\bibitem [{\citenamefont {Degen}\ \emph {et~al.}(2017)\citenamefont {Degen},
  \citenamefont {Reinhard},\ and\ \citenamefont
  {Cappellaro}}]{degen2017quantum}%
  \BibitemOpen
  \bibfield  {author} {\bibinfo {author} {\bibfnamefont {Christian~L}\
  \bibnamefont {Degen}}, \bibinfo {author} {\bibfnamefont {F}~\bibnamefont
  {Reinhard}}, \ and\ \bibinfo {author} {\bibfnamefont {P}~\bibnamefont
  {Cappellaro}},\ }\bibfield  {title} {\enquote {\bibinfo {title} {Quantum
  sensing},}\ }\href@noop {} {\bibfield  {journal} {\bibinfo  {journal}
  {Reviews of modern physics}\ }\textbf {\bibinfo {volume} {89}},\ \bibinfo
  {pages} {035002} (\bibinfo {year} {2017})}\BibitemShut {NoStop}%
\bibitem [{\citenamefont {Zayats}\ \emph {et~al.}(2005)\citenamefont {Zayats},
  \citenamefont {Smolyaninov},\ and\ \citenamefont
  {Maradudin}}]{zayats2005nano}%
  \BibitemOpen
  \bibfield  {author} {\bibinfo {author} {\bibfnamefont {Anatoly~V}\
  \bibnamefont {Zayats}}, \bibinfo {author} {\bibfnamefont {Igor~I}\
  \bibnamefont {Smolyaninov}}, \ and\ \bibinfo {author} {\bibfnamefont
  {Alexei~A}\ \bibnamefont {Maradudin}},\ }\bibfield  {title} {\enquote
  {\bibinfo {title} {Nano-optics of surface plasmon polaritons},}\ }\href@noop
  {} {\bibfield  {journal} {\bibinfo  {journal} {Physics reports}\ }\textbf
  {\bibinfo {volume} {408}},\ \bibinfo {pages} {131--314} (\bibinfo {year}
  {2005})}\BibitemShut {NoStop}%
\bibitem [{\citenamefont {Hong}\ \emph {et~al.}(2017)\citenamefont {Hong},
  \citenamefont {Li}, \citenamefont {Yang},\ and\ \citenamefont
  {Sengupta}}]{hong2017fully}%
  \BibitemOpen
  \bibfield  {author} {\bibinfo {author} {\bibfnamefont {Lingyu}\ \bibnamefont
  {Hong}}, \bibinfo {author} {\bibfnamefont {Hao}\ \bibnamefont {Li}}, \bibinfo
  {author} {\bibfnamefont {Haw}\ \bibnamefont {Yang}}, \ and\ \bibinfo {author}
  {\bibfnamefont {Kaushik}\ \bibnamefont {Sengupta}},\ }\bibfield  {title}
  {\enquote {\bibinfo {title} {Fully integrated fluorescence biosensors on-chip
  employing multi-functional nanoplasmonic optical structures in cmos},}\
  }\href@noop {} {\bibfield  {journal} {\bibinfo  {journal} {IEEE Journal of
  Solid-State Circuits}\ }\textbf {\bibinfo {volume} {52}},\ \bibinfo {pages}
  {2388--2406} (\bibinfo {year} {2017})}\BibitemShut {NoStop}%
\bibitem [{\citenamefont {Murari}\ \emph {et~al.}(2009)\citenamefont {Murari},
  \citenamefont {Etienne-Cummings}, \citenamefont {Thakor},\ and\ \citenamefont
  {Cauwenberghs}}]{murari2009photodiode}%
  \BibitemOpen
  \bibfield  {author} {\bibinfo {author} {\bibfnamefont {Kartikeya}\
  \bibnamefont {Murari}}, \bibinfo {author} {\bibfnamefont {Ralph}\
  \bibnamefont {Etienne-Cummings}}, \bibinfo {author} {\bibfnamefont {Nitish}\
  \bibnamefont {Thakor}}, \ and\ \bibinfo {author} {\bibfnamefont {Gert}\
  \bibnamefont {Cauwenberghs}},\ }\bibfield  {title} {\enquote {\bibinfo
  {title} {Which photodiode to use: a comparison of cmos-compatible
  structures},}\ }\href@noop {} {\bibfield  {journal} {\bibinfo  {journal}
  {IEEE sensors journal}\ }\textbf {\bibinfo {volume} {9}},\ \bibinfo {pages}
  {752--760} (\bibinfo {year} {2009})}\BibitemShut {NoStop}%
\bibitem [{\citenamefont {Yue}\ and\ \citenamefont {Wong}(1998)}]{yue1998chip}%
  \BibitemOpen
  \bibfield  {author} {\bibinfo {author} {\bibfnamefont {C~Patrick}\
  \bibnamefont {Yue}}\ and\ \bibinfo {author} {\bibfnamefont {S~Simon}\
  \bibnamefont {Wong}},\ }\bibfield  {title} {\enquote {\bibinfo {title}
  {On-chip spiral inductors with patterned ground shields for si-based rf
  ics},}\ }\href@noop {} {\bibfield  {journal} {\bibinfo  {journal} {IEEE
  Journal of solid-state circuits}\ }\textbf {\bibinfo {volume} {33}},\
  \bibinfo {pages} {743--752} (\bibinfo {year} {1998})}\BibitemShut {NoStop}%
\bibitem [{\citenamefont {Peng}\ and\ \citenamefont
  {Morris}(1996)}]{peng1996resonant}%
  \BibitemOpen
  \bibfield  {author} {\bibinfo {author} {\bibfnamefont {Song}\ \bibnamefont
  {Peng}}\ and\ \bibinfo {author} {\bibfnamefont {G~Michael}\ \bibnamefont
  {Morris}},\ }\bibfield  {title} {\enquote {\bibinfo {title} {Resonant
  scattering from two-dimensional gratings},}\ }\href@noop {} {\bibfield
  {journal} {\bibinfo  {journal} {JOSA A}\ }\textbf {\bibinfo {volume} {13}},\
  \bibinfo {pages} {993--1005} (\bibinfo {year} {1996})}\BibitemShut {NoStop}%
\bibitem [{\citenamefont {Charbon}\ \emph {et~al.}(2017)\citenamefont
  {Charbon}, \citenamefont {Sebastiano}, \citenamefont {Babaie}, \citenamefont
  {Vladimirescu}, \citenamefont {Shahmohammadi}, \citenamefont {Staszewski},
  \citenamefont {Homulle}, \citenamefont {Patra}, \citenamefont {Van~Dijk},
  \citenamefont {Incandela} \emph {et~al.}}]{charbon201715}%
  \BibitemOpen
  \bibfield  {author} {\bibinfo {author} {\bibfnamefont {Edoardo}\ \bibnamefont
  {Charbon}}, \bibinfo {author} {\bibfnamefont {Fabio}\ \bibnamefont
  {Sebastiano}}, \bibinfo {author} {\bibfnamefont {Masoud}\ \bibnamefont
  {Babaie}}, \bibinfo {author} {\bibfnamefont {Andrei}\ \bibnamefont
  {Vladimirescu}}, \bibinfo {author} {\bibfnamefont {Mina}\ \bibnamefont
  {Shahmohammadi}}, \bibinfo {author} {\bibfnamefont {Robert~Bogdan}\
  \bibnamefont {Staszewski}}, \bibinfo {author} {\bibfnamefont {Harald~AR}\
  \bibnamefont {Homulle}}, \bibinfo {author} {\bibfnamefont {Bishnu}\
  \bibnamefont {Patra}}, \bibinfo {author} {\bibfnamefont {Jeroen~PG}\
  \bibnamefont {Van~Dijk}}, \bibinfo {author} {\bibfnamefont {Rosario~M}\
  \bibnamefont {Incandela}},  \emph {et~al.},\ }\bibfield  {title} {\enquote
  {\bibinfo {title} {15.5 cryo-cmos circuits and systems for scalable quantum
  computing},}\ }in\ \href@noop {} {\emph {\bibinfo {booktitle} {Solid-State
  Circuits Conference (ISSCC), 2017 IEEE International}}}\ (\bibinfo
  {organization} {Ieee},\ \bibinfo {year} {2017})\ pp.\ \bibinfo {pages}
  {264--265}\BibitemShut {NoStop}%
\bibitem [{\citenamefont {Yao}\ \emph {et~al.}(2012)\citenamefont {Yao},
  \citenamefont {Jiang}, \citenamefont {Gorshkov}, \citenamefont {Maurer},
  \citenamefont {Giedke}, \citenamefont {Cirac},\ and\ \citenamefont
  {Lukin}}]{yao2012scalable}%
  \BibitemOpen
  \bibfield  {author} {\bibinfo {author} {\bibfnamefont {Norman~Y}\
  \bibnamefont {Yao}}, \bibinfo {author} {\bibfnamefont {Liang}\ \bibnamefont
  {Jiang}}, \bibinfo {author} {\bibfnamefont {Alexey~V}\ \bibnamefont
  {Gorshkov}}, \bibinfo {author} {\bibfnamefont {Peter~C}\ \bibnamefont
  {Maurer}}, \bibinfo {author} {\bibfnamefont {Geza}\ \bibnamefont {Giedke}},
  \bibinfo {author} {\bibfnamefont {J~Ignacio}\ \bibnamefont {Cirac}}, \ and\
  \bibinfo {author} {\bibfnamefont {Mikhail~D}\ \bibnamefont {Lukin}},\
  }\bibfield  {title} {\enquote {\bibinfo {title} {Scalable architecture for a
  room temperature solid-state quantum information processor},}\ }\href@noop {}
  {\bibfield  {journal} {\bibinfo  {journal} {Nature communications}\ }\textbf
  {\bibinfo {volume} {3}},\ \bibinfo {pages} {800} (\bibinfo {year}
  {2012})}\BibitemShut {NoStop}%
\bibitem [{\citenamefont {Veldhorst}\ \emph {et~al.}(2017)\citenamefont
  {Veldhorst}, \citenamefont {Eenink}, \citenamefont {Yang},\ and\
  \citenamefont {Dzurak}}]{veldhorst2017silicon}%
  \BibitemOpen
  \bibfield  {author} {\bibinfo {author} {\bibfnamefont {M}~\bibnamefont
  {Veldhorst}}, \bibinfo {author} {\bibfnamefont {HGJ}\ \bibnamefont {Eenink}},
  \bibinfo {author} {\bibfnamefont {CH}~\bibnamefont {Yang}}, \ and\ \bibinfo
  {author} {\bibfnamefont {AS}~\bibnamefont {Dzurak}},\ }\bibfield  {title}
  {\enquote {\bibinfo {title} {Silicon cmos architecture for a spin-based
  quantum computer},}\ }\href@noop {} {\bibfield  {journal} {\bibinfo
  {journal} {Nature communications}\ }\textbf {\bibinfo {volume} {8}},\
  \bibinfo {pages} {1766} (\bibinfo {year} {2017})}\BibitemShut {NoStop}%
\bibitem [{\citenamefont {Patra}\ \emph {et~al.}(2018)\citenamefont {Patra},
  \citenamefont {Incandela}, \citenamefont {Van~Dijk}, \citenamefont {Homulle},
  \citenamefont {Song}, \citenamefont {Shahmohammadi}, \citenamefont
  {Staszewski}, \citenamefont {Vladimirescu}, \citenamefont {Babaie},
  \citenamefont {Sebastiano} \emph {et~al.}}]{patra2018cryo}%
  \BibitemOpen
  \bibfield  {author} {\bibinfo {author} {\bibfnamefont {Bishnu}\ \bibnamefont
  {Patra}}, \bibinfo {author} {\bibfnamefont {Rosario~M}\ \bibnamefont
  {Incandela}}, \bibinfo {author} {\bibfnamefont {Jeroen~PG}\ \bibnamefont
  {Van~Dijk}}, \bibinfo {author} {\bibfnamefont {Harald~AR}\ \bibnamefont
  {Homulle}}, \bibinfo {author} {\bibfnamefont {Lin}\ \bibnamefont {Song}},
  \bibinfo {author} {\bibfnamefont {Mina}\ \bibnamefont {Shahmohammadi}},
  \bibinfo {author} {\bibfnamefont {Robert~Bogdan}\ \bibnamefont {Staszewski}},
  \bibinfo {author} {\bibfnamefont {Andrei}\ \bibnamefont {Vladimirescu}},
  \bibinfo {author} {\bibfnamefont {Masoud}\ \bibnamefont {Babaie}}, \bibinfo
  {author} {\bibfnamefont {Fabio}\ \bibnamefont {Sebastiano}},  \emph
  {et~al.},\ }\bibfield  {title} {\enquote {\bibinfo {title} {Cryo-cmos
  circuits and systems for quantum computing applications},}\ }\href@noop {}
  {\bibfield  {journal} {\bibinfo  {journal} {IEEE Journal of Solid-State
  Circuits}\ } (\bibinfo {year} {2018})}\BibitemShut {NoStop}%
\bibitem [{\citenamefont {Giovannetti}\ \emph {et~al.}(2011)\citenamefont
  {Giovannetti}, \citenamefont {Lloyd},\ and\ \citenamefont
  {Maccone}}]{giovannetti2011advances}%
  \BibitemOpen
  \bibfield  {author} {\bibinfo {author} {\bibfnamefont {Vittorio}\
  \bibnamefont {Giovannetti}}, \bibinfo {author} {\bibfnamefont {Seth}\
  \bibnamefont {Lloyd}}, \ and\ \bibinfo {author} {\bibfnamefont {Lorenzo}\
  \bibnamefont {Maccone}},\ }\bibfield  {title} {\enquote {\bibinfo {title}
  {Advances in quantum metrology},}\ }\href@noop {} {\bibfield  {journal}
  {\bibinfo  {journal} {Nature photonics}\ }\textbf {\bibinfo {volume} {5}},\
  \bibinfo {pages} {222} (\bibinfo {year} {2011})}\BibitemShut {NoStop}%
\bibitem [{\citenamefont {Unden}\ \emph {et~al.}(2016)\citenamefont {Unden},
  \citenamefont {Balasubramanian}, \citenamefont {Louzon}, \citenamefont
  {Vinkler}, \citenamefont {Plenio}, \citenamefont {Markham}, \citenamefont
  {Twitchen}, \citenamefont {Stacey}, \citenamefont {Lovchinsky}, \citenamefont
  {Sushkov} \emph {et~al.}}]{unden2016quantum}%
  \BibitemOpen
  \bibfield  {author} {\bibinfo {author} {\bibfnamefont {Thomas}\ \bibnamefont
  {Unden}}, \bibinfo {author} {\bibfnamefont {Priya}\ \bibnamefont
  {Balasubramanian}}, \bibinfo {author} {\bibfnamefont {Daniel}\ \bibnamefont
  {Louzon}}, \bibinfo {author} {\bibfnamefont {Yuval}\ \bibnamefont {Vinkler}},
  \bibinfo {author} {\bibfnamefont {Martin~B}\ \bibnamefont {Plenio}}, \bibinfo
  {author} {\bibfnamefont {Matthew}\ \bibnamefont {Markham}}, \bibinfo {author}
  {\bibfnamefont {Daniel}\ \bibnamefont {Twitchen}}, \bibinfo {author}
  {\bibfnamefont {Alastair}\ \bibnamefont {Stacey}}, \bibinfo {author}
  {\bibfnamefont {Igor}\ \bibnamefont {Lovchinsky}}, \bibinfo {author}
  {\bibfnamefont {Alexander~O}\ \bibnamefont {Sushkov}},  \emph {et~al.},\
  }\bibfield  {title} {\enquote {\bibinfo {title} {Quantum metrology enhanced
  by repetitive quantum error correction},}\ }\href@noop {} {\bibfield
  {journal} {\bibinfo  {journal} {Physical review letters}\ }\textbf {\bibinfo
  {volume} {116}},\ \bibinfo {pages} {230502} (\bibinfo {year}
  {2016})}\BibitemShut {NoStop}%
\bibitem [{\citenamefont {Bernien}\ \emph {et~al.}(2013)\citenamefont
  {Bernien}, \citenamefont {Hensen}, \citenamefont {Pfaff}, \citenamefont
  {Koolstra}, \citenamefont {Blok}, \citenamefont {Robledo}, \citenamefont
  {Taminiau}, \citenamefont {Markham}, \citenamefont {Twitchen}, \citenamefont
  {Childress} \emph {et~al.}}]{bernien2013heralded}%
  \BibitemOpen
  \bibfield  {author} {\bibinfo {author} {\bibfnamefont {Hannes}\ \bibnamefont
  {Bernien}}, \bibinfo {author} {\bibfnamefont {Bas}\ \bibnamefont {Hensen}},
  \bibinfo {author} {\bibfnamefont {Wolfgang}\ \bibnamefont {Pfaff}}, \bibinfo
  {author} {\bibfnamefont {Gerwin}\ \bibnamefont {Koolstra}}, \bibinfo {author}
  {\bibfnamefont {MS}~\bibnamefont {Blok}}, \bibinfo {author} {\bibfnamefont
  {Lucio}\ \bibnamefont {Robledo}}, \bibinfo {author} {\bibfnamefont
  {TH}~\bibnamefont {Taminiau}}, \bibinfo {author} {\bibfnamefont {Matthew}\
  \bibnamefont {Markham}}, \bibinfo {author} {\bibfnamefont {DJ}~\bibnamefont
  {Twitchen}}, \bibinfo {author} {\bibfnamefont {Lilian}\ \bibnamefont
  {Childress}},  \emph {et~al.},\ }\bibfield  {title} {\enquote {\bibinfo
  {title} {Heralded entanglement between solid-state qubits separated by three
  metres},}\ }\href@noop {} {\bibfield  {journal} {\bibinfo  {journal}
  {Nature}\ }\textbf {\bibinfo {volume} {497}},\ \bibinfo {pages} {86}
  (\bibinfo {year} {2013})}\BibitemShut {NoStop}%
\bibitem [{\citenamefont {Pfaff}\ \emph {et~al.}(2014)\citenamefont {Pfaff},
  \citenamefont {Hensen}, \citenamefont {Bernien}, \citenamefont {van Dam},
  \citenamefont {Blok}, \citenamefont {Taminiau}, \citenamefont {Tiggelman},
  \citenamefont {Schouten}, \citenamefont {Markham}, \citenamefont {Twitchen}
  \emph {et~al.}}]{pfaff2014unconditional}%
  \BibitemOpen
  \bibfield  {author} {\bibinfo {author} {\bibfnamefont {Wolfgang}\
  \bibnamefont {Pfaff}}, \bibinfo {author} {\bibfnamefont {BJ}~\bibnamefont
  {Hensen}}, \bibinfo {author} {\bibfnamefont {Hannes}\ \bibnamefont
  {Bernien}}, \bibinfo {author} {\bibfnamefont {Suzanne~B}\ \bibnamefont {van
  Dam}}, \bibinfo {author} {\bibfnamefont {Machiel~S}\ \bibnamefont {Blok}},
  \bibinfo {author} {\bibfnamefont {Tim~H}\ \bibnamefont {Taminiau}}, \bibinfo
  {author} {\bibfnamefont {Marijn~J}\ \bibnamefont {Tiggelman}}, \bibinfo
  {author} {\bibfnamefont {Raymond~N}\ \bibnamefont {Schouten}}, \bibinfo
  {author} {\bibfnamefont {Matthew}\ \bibnamefont {Markham}}, \bibinfo {author}
  {\bibfnamefont {Daniel~J}\ \bibnamefont {Twitchen}},  \emph {et~al.},\
  }\bibfield  {title} {\enquote {\bibinfo {title} {Unconditional quantum
  teleportation between distant solid-state quantum bits},}\ }\href@noop {}
  {\bibfield  {journal} {\bibinfo  {journal} {Science}\ }\textbf {\bibinfo
  {volume} {345}},\ \bibinfo {pages} {532--535} (\bibinfo {year}
  {2014})}\BibitemShut {NoStop}%
\bibitem [{\citenamefont {Humphreys}\ \emph {et~al.}(2018)\citenamefont
  {Humphreys}, \citenamefont {Kalb}, \citenamefont {Morits}, \citenamefont
  {Schouten}, \citenamefont {Vermeulen}, \citenamefont {Twitchen},
  \citenamefont {Markham},\ and\ \citenamefont
  {Hanson}}]{humphreys2018deterministic}%
  \BibitemOpen
  \bibfield  {author} {\bibinfo {author} {\bibfnamefont {Peter~C}\ \bibnamefont
  {Humphreys}}, \bibinfo {author} {\bibfnamefont {Norbert}\ \bibnamefont
  {Kalb}}, \bibinfo {author} {\bibfnamefont {Jaco~PJ}\ \bibnamefont {Morits}},
  \bibinfo {author} {\bibfnamefont {Raymond~N}\ \bibnamefont {Schouten}},
  \bibinfo {author} {\bibfnamefont {Raymond~FL}\ \bibnamefont {Vermeulen}},
  \bibinfo {author} {\bibfnamefont {Daniel~J}\ \bibnamefont {Twitchen}},
  \bibinfo {author} {\bibfnamefont {Matthew}\ \bibnamefont {Markham}}, \ and\
  \bibinfo {author} {\bibfnamefont {Ronald}\ \bibnamefont {Hanson}},\
  }\bibfield  {title} {\enquote {\bibinfo {title} {Deterministic delivery of
  remote entanglement on a quantum network},}\ }\href@noop {} {\bibfield
  {journal} {\bibinfo  {journal} {Nature}\ }\textbf {\bibinfo {volume} {558}},\
  \bibinfo {pages} {268} (\bibinfo {year} {2018})}\BibitemShut {NoStop}%
\end{thebibliography}
\end{document}